\documentclass[aps,twocolumn,showpacs,preprintnumbers,amsmath,amssymb,nofootinbib,superscriptaddress,showkeys]{revtex4}

\usepackage{hyperref}
\usepackage{float}
\usepackage{epsfig}
\usepackage{graphicx}
\usepackage{mathptmx}      
\usepackage{latexsym}
\usepackage{longtable}
\usepackage{dcolumn}

\begin{document}

\title{Coarse grained NN potential with  Chiral Two Pion Exchange}~\thanks{Supported by Spanish DGI (grant
  FIS2011-24149) and Junta de Andaluc{\'{\i}a} (grant FQM225).
  R.N.P. is supported by a Mexican CONACYT grant.}

\author{R. Navarro P\'erez}\email{rnavarrop@ugr.es}
\affiliation{Departamento de F\'{\i}sica At\'omica, Molecular y
  Nuclear \\ and Instituto Carlos I de F{\'\i}sica Te\'orica y
  Computacional \\ Universidad de Granada, E-18071 Granada, Spain.}
\author{J.E. Amaro}\email{amaro@ugr.es} \affiliation{Departamento de
  F\'{\i}sica At\'omica, Molecular y Nuclear \\ and Instituto Carlos I
  de F{\'\i}sica Te\'orica y Computacional \\ Universidad de Granada,
  E-18071 Granada, Spain.}  \author{E. Ruiz
  Arriola}\email{earriola@ugr.es} \affiliation{Departamento de
  F\'{\i}sica At\'omica, Molecular y Nuclear \\ and Instituto Carlos I
  de F{\'\i}sica Te\'orica y Computacional \\ Universidad de Granada,
  E-18071 Granada, Spain.}

\date{\today}

\begin{abstract} 
\rule{0ex}{3ex} We determine the chiral constants of the
Nucleon-Nucleon Two Pion Exchange potential deduced from Chiral
Perturbation Theory.  By using a coarse grained representation of the
short distance interactions with $30$ parameters, the Partial Wave
Analysis fit gives $\chi^2/\nu = 1.08 $ to a mutually consistent set of
6713 data previously built from all published
proton-proton and neutron proton scattering data from 1950 till 2013
with LAB energy below 350 MeV. We obtain $(c_1, c_3, c_4)=(-0.41 \pm
1.08,-4.66 \pm 0.60, 4.31 \pm 0.17)\,  {\rm GeV}^{-1} $ with an almost
$100\%$ anti-correlation between $c_1$ and $c_3$. We also provide the
errors in the short distance parameters and propagate them to the
deuteron properties and low partial waves phase shifts.
\end{abstract}

\maketitle

\section{Introduction}
\label{sec:intro}

The modern chiral theory of Nuclear Forces era started in 1990 when Weinberg
suggested~\cite{Weinberg:1990rz} using Effective Field Theory in
conjunction with Chiral Symmetry to derive in a systematic and
model-independent way the forces between many nucleons complying with
the symmetries of Quantum Chromodynamics (QCD). The Chiral
Perturbation Theory (ChPT) predicts an increasing suppression of
n-body forces at long distances and hence was further
elaborated~\cite{Ordonez:1992xp} and confronted to NN data soon
thereafter~\cite{Ordonez:1995rz}. This requires the introduction of
counterterms encoding the {\it unknown} short distance piece of the
interaction and which are not directly constrained by chiral symmetry
(see e.g. \cite{Epelbaum:2008ga,Machleidt:2011zz} for reviews).

While One Pion Exchange (OPE) is a quite universal feature of most
phenomenological NN interactions and a simple consequence of the meson
exchange picture, Chiral Two Pion Exchange ($\chi$TPE) arises as a
consequence of the spontaneous breakdown of chiral symmetry and 
the chiral constants $c_1,c_3$ and $c_4$ appearing in $\pi N$
scattering at low energies emerge at the Next-to-next-to-leading order
(N2LO) in the chiral expansion of the NN
force~\cite{Kaiser:1997mw}. Because the NN interaction is a basic
building block in Nuclear Physics, the consistency of both
determinations is a necessary and important condition for the
verification of this upgraded view of Nuclear Physics. A comparative
overview of different $\pi N$ and $NN$ determinations up to 2005 is
presented in Ref.~\cite{Epelbaum:2005pn}.

Our purpose is to extract $c_1$,$c_3$ and $c_4$ from a Partial Wave
Analysis (PWA) of the 8124 published proton-proton and neutron-proton
scattering data collected from 1950 till 2013 and using the NN chiral
potential up to N2LO in the Weinberg counting~\cite{Kaiser:1997mw}. We
stress that we are {\it not} making a ChPT calculation which would
only apply below energies sensing the $3\pi$-exchange left cut,
$E_{\rm LAB}= (2/M_N)(3m_\pi/2)^2 \lesssim 100 {\rm MeV}$. We rather
determine the long distance tail of the potential constraining the
short distance interaction with higher energies. We remind that,
according to well known statistical principles, it is essential to
validate the fit to the data with a $\chi^2$ per degree of freedom
$\chi^2 /\nu \sim 1$ with $\nu= N_{\rm Data}-N_{\rm par}$ before
errors in fitting parameters can be determined. 

Much of the present understanding of NN interactions has profited
inmensely from the long term in-depth studies of the Nijmegen group,
which culminated with the concept of high quality interactions,
i.e. with $\chi^2 /\nu \sim
1$~\cite{Stoks:1993tb,Stoks:1994wp}. Subsequent analyses have been
built upon these works by incorporating new data and potential
forms~\cite{Wiringa:1994wb,Machleidt:2000ge,Gross:2008ps} including
the chiral TPE analysis of the Nijmegen
group~\cite{Rentmeester:1999vw,Rentmeester:2003mf}. In our most recent
work~\cite{Perez:2013mwa,Perez:2013jpa} a refined rejection criterium
was applied and a large number of data published since the original
Nijmegen PWA below pion producion
threshold~\cite{Stoks:1993tb,Stoks:1994wp} have been added to the
database, almost doubling the total number.  The present work
represents an upgrade of the chiral
TPE-PWA~\cite{Rentmeester:1999vw,Rentmeester:2003mf} with this new
data set keeping identical the long range part of the interaction, in
particular the OPE and TPE part as well as the electromagnetic effects,
but using the computationally convenient $\delta$-shell
representation~\cite{Perez:2013mwa,Perez:2013jpa} for the unknown
short range contribution to the NN-potential.

The paper is organized as follows. In section~\ref{sec:NN-coarse} we
describe the main new issues considered in our analysis. Details of
the fit involving $\chi$TPE are discussed in
Section~\ref{sec:fit-tpe}. After that, in Section~\ref{sec:error-an},
we discuss the errors analysis of our fits. Using the covariance
matrix obtained from our analysis of the data, we are in a position to
propagate uncertainties and list np and pp phases with statistical
errors based on $\chi$TPE potentials in
Section~\ref{sec:prop}. Finally, in Section~\ref{sec:conc} we come to
our main conclusions.

\section{NN data and Coarse grained potentials}
\label{sec:NN-coarse}

The large body of published data is not fully consistent, as
recognized by earlier high quality
fits~\cite{Stoks:1993tb,Stoks:1994wp,Wiringa:1994wb,Machleidt:2000ge,Gross:2008ps},
i.e. having $\chi^2/\nu \lesssim 1 $. The problem was handled by using
a rejection criterion at the $3\sigma$ confidence level. In
Ref.~\cite{Perez:2013mwa,Perez:2013jpa} we use a procedure suggested
by Gross and Stadler~\cite{Gross:2008ps} which essentially provides a
self-consistent way of analyzing the tension among all the data and
$3\sigma$-rejecting mutually inconsistent data. This is done by using
a charge dependent OPE potential plus electromagnetic effects such as
vacuum polarization, magnetic moments interaction, etc. above a
cut-off radius of $r_c=3 {\rm fm}$ (see Ref.~\cite{Perez:2013jpa} for
a recollection of formulas). The short range part is most conveniently
parameterized following Aviles~\cite{Aviles:1973ee} as a sum of Dirac
delta-shells located at equidistant points below $r_c$ and separated
by $\Delta r = 0.6 {\rm fm}$ (see also
\cite{Entem:2007jg,NavarroPerez:2011fm,NavarroPerez:2012qf,Perez:2013za,Perez:2013cza}
for further details and applications). The short range NN interaction
can be written as a sum of delta-shells, so that the total potential
reads
\begin{eqnarray}
   V(r) &=& \sum_{n=1}^{21} O_n \left[\sum_{i=1}^N V_{i,n} \Delta r_i \delta(r-r_i) \right] \nonumber \\ &+& V_{\rm long} (r) \theta(r-r_c),
\label{eq:potential}
\end{eqnarray}
where $O_n$ are the set of operators in the AV18
basis~\cite{Wiringa:1994wb}, $r_i \le r_c$ are a discrete set of
$N$-radii, $\Delta r_i = r_{i+1}-r_i $ and $V_{i,n}$ are unknown
coefficients to be determined from data.  The $r > r_c$ piece ,
$V_{\rm long}(r)$ contains a Charge-Dependent (CD) One pion exchange
(OPE) and electromagnetic (EM) corrections which are kept fixed
throughout
\begin{eqnarray}
V_{\rm long}(r) = V_{\rm OPE}(r) + V_{\rm em}(r) \, . 
\end{eqnarray}
The form of the complete potential includes an operator basis
extending the AV18 potential~\cite{Wiringa:1994wb} and specified in
Ref.~\cite{Perez:2013mwa,Perez:2013jpa} but the statistical analysis
is carried out more effectively in terms of some low and independent
partial waves contributions to the potential from which all other
higher partial waves are consistently deduced (see
Ref.~\cite{Perez:2013mwa,Perez:2013jpa}). The PWA allows to accept
$N_{\rm accept}=6713$ data with a $\chi^2/\nu = 1.04$. The present
work uses this fixed database which is extensively described in
Ref.~\cite{Perez:2013mwa,Perez:2013jpa}, and the {\it same} long-range
potentials.

\section{Fit of  Two Pion Exchange potential}
\label{sec:fit-tpe}

In this work we keep the OPE piece with the recommended value
$f^2=0.075$~\cite{Stoks:1992ja,deSwart:1997ep} as we did in
Refs.~\cite{Perez:2013mwa,Perez:2013jpa} and add the $\chi$TPE
potential~\cite{Kaiser:1997mw} to the long range piece,
\begin{eqnarray}
V_{\rm long}(r) = V_{\rm \chi TPE} (r)+ 
V_{\rm OPE}(r) + V_{\rm em}(r) \, . 
\end{eqnarray}
We also modify the cut-off radius $r_c$ to be to be determined from a
fit to the data. Namely, we take the values $r_c=3,2.4,1.8$fm. This
reduces the number of delta-shells and hence the number of short
distance parameters $\lambda_{i,n}$. The three chiral constants
$c_1,c_3$ and $c_4$ of the $\chi$TPE potential will be additional
parameters of the fit.  Since we aim at a determination of
uncertainties in these parameters we can only do so provided the fit
is acceptable, i.e. $\chi^2/\nu \sim 1 $. The quality of our fits
regarding the influence of TPE in the description of the data can be
judged by analyzing three different schemes which are displayed in
tables \ref{tab:1},\ref{tab:2} and \ref{tab:3}.

In table \ref{tab:1} we show the $\chi^2$ values corresponding to a
direct fit to all the data without rejecting any of the published
experimental results gathered from 1950 until 2013. As we see, the
large $ \chi^2$-values correspond to an unacceptable fit and hence
prevent error determination and propagation.  In table \ref{tab:2} we
show the $\chi^2$ values corresponding to a dynamical data base fit to
all the data subjected to the $3\sigma$
criterion~\cite{Stoks:1993tb,Stoks:1994wp,Wiringa:1994wb,Machleidt:2000ge,Gross:2008ps},
so that the selection of the data depends on the description of the
long range interaction which in our case is $\chi$TPE and on the value
of the cut-off radius $r_c$.  As we see, there is a reduction on the
$\chi^2$ value but the number of rejected data differ among each
other.  The data rejection triggered by the $\chi$TPE potential does
not correspond to eliminate mutually inconsistent data, but rather to
shape the data base to better comply to the chiral theory, and in our
view represents a bias which definitely induces a systematic error in
the analysis. Finally, in table \ref{tab:3} we use the fixed and
consistent data from the OPE $r_c=3 {\rm fm}$ analysis based on the
improved 3$\sigma$ criterion of Gross and Stadler~\cite{Gross:2008ps}
carried out in practice in our recent work~\cite{Perez:2013za}. In
this case, an acceptable $\chi^2=1.1$ with 30 parameters allows to
determine and propagate errors.

A comprehensive overview of several high quality analyzes up to
$E_{\rm LAB} \le 350 {\rm MeV}$ is presented in Table~\ref{chi2}. This
includes PWA93~\cite{Stoks:1993tb}, Nijm I~\cite{Stoks:1994wp}, Nijm
II~\cite{Stoks:1994wp}, Reid93~\cite{Stoks:1994wp} ,
AV18~\cite{Wiringa:1994wb}, CD-Bonn~\cite{Machleidt:2000ge} , WJC1 and
WJC2~\cite{Gross:2008ps}, PWApp-TPE~\cite{Rentmeester:1999vw} and
PWANN-TPE~\cite{Rentmeester:2003mf} (here $E_{\rm LAB} \le 500 {\rm
  MeV}$) as well as our recent $\delta$shell-OPE
fit~\cite{Perez:2013mwa}. As one sees the quality of the fit depends
both on the number of parameters as well as the total number of
analyzed data. 

\begin{table*}[t]
\caption{Complete  NN database from PWA without rejection. $N_{\rm Data}=8124$.}
\centering
\label{tab:1}       
	\begin{tabular}{llllllllll}
\hline\noalign{\smallskip}
      $r_c$ [fm]  &          & 1.8 &   &  & 2.4  & & & 3.0 &  \\ 
             &    & $N_{\rm p}$ & $\chi^2/\nu$ &   & $N_{\rm p}$ & $\chi^2/\nu$ &  & $N_{\rm p}$ & $\chi^2/\nu$ \\
      OPE  &   & 31   & {1.80}     &  & 39   & 1.56     &  & {46}   & {1.54} \\
      TPE(NLO) &   & 31   & {1.72}     &  & 38   & 1.56     &  & 46   & 1.52 \\
      TPE(N2LO) & \ \ \ \ \ \ \ \ \  & {30+3} & {1.60}  & \ \ \ \ \ \ \ \ \  & 38+3 & 1.56     & \ \ \ \ \ \ \ \ \ & 46+3 & 1.52
	\end{tabular} 
\end{table*}
\begin{table*}[t]
\caption{$3\sigma$-selected  NN database from potential analysis.}
\centering
\label{tab:2}       
	\begin{tabular}{llllllllll}
\hline\noalign{\smallskip}
      $r_c$ [fm]  &          & 1.8 &   &  & 2.4  & & & 3.0 &  \\ 
             & $N_{\rm accept}$   & $N_{\rm par}$ & $\chi^2/\nu$ &   $N_{\rm accept}$  & $N_{\rm par}$ & $\chi^2/\nu$ &  $N_{\rm accept}$ & $N_{\rm par}$ & $\chi^2/\nu$ \\
      OPE  & {5766}  & 31   & 1.10     &    6363 & 39   & 1.09     &  {6438} & {46}   & 1.06 \\
      TPE(NLO) & {5841}  & 31   & 1.10     &    6432 & 38   & 1.10     &    6423 & 46   & 1.06 \\
      TPE(N2LO) & {6220}  & {30+3} & 1.07     &    6439 & 38+3 & 1.10     &    6422 & 46+3 & 1.06
	\end{tabular}
\end{table*}
\begin{table*}[t]
\caption{Consistent NN database from the improved $3\sigma$-criterion. $N_{\rm Data}= N_{\rm accept}^{({\rm OPE},r_c=3 {\rm fm})}=6713$.}
\centering
\label{tab:3}       
	\begin{tabular}{lllllll}
\hline\noalign{\smallskip}
      $r_c$ [fm]  & 1.8    &       & 2.4 & &  3.0 & \\
        & $N_{\rm par}$ & $\chi^2/\nu$   & $N_{\rm par}$ & $\chi^2/\nu$  & $N_{\rm par}$ & $\chi^2/\nu$ \\
      OPE  &  31   & {1.37}   & 39   & 1.09    & {46}   & {1.06} \\
      TPE(NLO)  & 31   & {1.26}      & 38   & 1.08    & 46   & 1.06 \\
      TPE(NNLO) & {30+3} & {1.10}     & 38+3 & 1.08     & 46+3 & 1.06
	\end{tabular}
\end{table*}

\section{Error analysis with TPE potential}
\label{sec:error-an}

As already mentioned, the inclusion of the $\chi$TPE
potential~\cite{Kaiser:1997mw} allows to describe the interaction in
the region below $3 {\rm fm}$ and reduces the cut-off radius down to
$r_c=1.8 {\rm fm}$, before sensing nucleon finite size effects (see
e.g. the discussion in Ref.~\cite{Perez:2013cza}).  Thus, some of the
delta-shells which generally coarse grain the interaction are removed
in favour of an underlying and explicit chiral representation.  As in
our previous PWA using OPE~\cite{Perez:2013mwa,Perez:2013jpa} we
impose the np and pp contributions to be identical in all isovector
partial waves except the $^1S_0$. This yields $\chi^2/\nu =1.08$, a
slightly higher value than with our OPE PWA, but improving over the
much used AV18 potential where $\chi^2/\nu=1.09$~\cite{Wiringa:1994wb}
where the number of data was about $60\%$ less than in the present 
analysis.  The most recent study based on the covariant spectator
model~\cite{Gross:2008ps} where only $np$ was considered (see Table
\ref{chi2}).

The resulting short distance parameters and
their errors are presented in table~\ref{tab:Fits}.  The first line
corresponds to a coarse graining of the {\it known} electromagnetic
part of the interaction as described
in~\cite{Perez:2013mwa,Perez:2013jpa} and, like there, they are fixed
throughout the fitting process. As we see only the innermost
$\lambda_1$ significantly differs by $25\%$ in the np and pp $^1S_0$
waves.

While this isospin violation 
prevents in our
view a sensible prediction for the nn $^1S_0$ scattering length based
solely on two body information (see
however~\cite{CalleCordon:2010sq}), it opens up an interesting
possibility regarding the inclusion of known isospin breaking effects
at the OPE and TPE level (see e.g. \cite{Miller:2006tv} for a
review). The small correction found in Ref.~\cite{vanKolck:1997fu}
requires an assumption on the regularization at short distances, which
in our approach is equivalent to treat the $^1S_0$ channel for np and
pp states as independent from each other.

\begin{table}[ttt]
	\centering
      \caption{Fitting delta-shell partial wave parameters
        $(\lambda_n)^{JS}_{l,l'} $ (in ${\rm fm}^{-1}$) with their
        errors for all states in the $JS$ channel. We take $N=3$
        equidistant points with $\Delta r = 0.6$fm. $-$ indicates that
        the corresponding fitting $(\lambda_n)^{JS}_{l,l'} =0$. In the
        first line we provide the central component of the delta
        shells corresponding to the EM effects below $r_c = 1.8 {\rm
          fm}$. These parameters remain fixed within the fitting
        process.}
      \label{tab:Fits}
	\begin{tabular*}{\columnwidth}{@{\extracolsep{\fill}}c c c c c}
            \hline
            \hline\noalign{\smallskip}
		Wave  & $\lambda_1$ & $\lambda_2$ & $\lambda_3$  \\

	   & ($r_1=0.6$fm) & ($r_2=1.2$fm) & ($r_3=1.8$fm)   \\

            \hline\noalign{\smallskip}
  $V_C[{\rm pp}]_{\rm EM}$  & 0.02069940 & 0.01871309 & 0.00460163 \\ \hline 
             $^1S_0[\rm np]$& 1.48(7)  & -0.86(1)  & -0.041(7) \\ 
             $^1S_0[\rm pp]$& 1.87(3)  & -0.875(5) & -0.045(3) \\ 
             $^3P_0$        & 2.318(3) &  0.400(7) & -0.093(3) \\ 
             $^1P_1$        &     $-$  &  1.09(1)  &      $-$  \\ 
             $^3P_1$        &     $-$  &  1.27(1)  &  0.008(3) \\ 
             $^3S_1$        & 1.16(2)  &      $-$  & -0.073(4) \\ 
             $\varepsilon_1$&     $-$  & -2.50(2)  & -0.097(4) \\ 
             $^3D_1$        &     $-$  &  2.03(6)  &      $-$  \\ 
             $^1D_2$        &     $-$  & -0.494(9) & -0.034(4) \\ 
             $^3D_2$        &     $-$  & -0.82(1)  & -0.148(5) \\ 
             $^3P_2$        &-0.953(4) & -0.233(4) & -0.034(2) \\ 
             $\varepsilon_2$&     $-$  &  0.85(2)  &  0.042(2) \\ 
             $^3F_2$        &     $-$  &  4.05(9)  & -0.110(4) \\ 
             $^1F_3$        &     $-$  &  1.7(1)   &      $-$  \\ 
             $^3D_3$        &     $-$  &  0.73(1)  &      $-$  \\ 
            \noalign{\smallskip}\hline
            \hline
	\end{tabular*}
\end{table}
 
\begin{figure*}[htb]
\centering
  \includegraphics[width=0.3\textwidth]{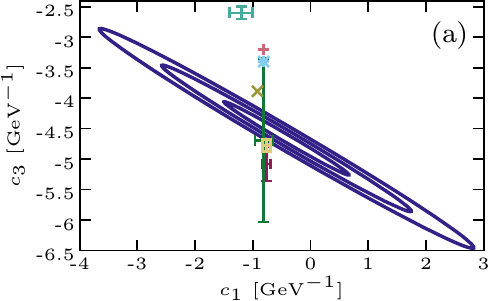}
  \includegraphics[width=0.3\textwidth]{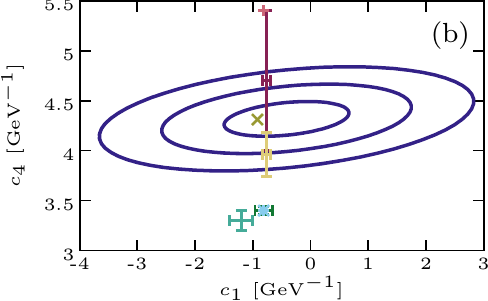}
  \includegraphics[width=0.3\textwidth]{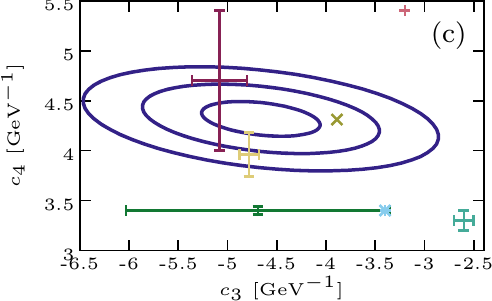}
\caption{Correlation ellipses for the chiral constants $c_1$ , $c_3$
  and $c_4$ appearing in the TPE potential with a cut-off radius of
  $r_c=1.8$fm from a PWA with the consistent database and with
  $\chi^2/\nu = 1.1$. The crosses represent the various determinations
  listed in table~\ref{tab:c1c3c4}.}
\label{fig:c1c3c4-ellipses}       
\end{figure*}

The correlation ellipses for $c_1$, $c_3$ and $c_4$ are presented for
$1\sigma$,$2\sigma$ and $3\sigma$ confidence levels in
Fig.~\ref{fig:c1c3c4-ellipses}. The numerical values can be looked up
in Table~\ref{tab:c1c3c4} and compared to other determinations based
on NN and $\pi N$ information (see e.g. Ref.~\cite{Alarcon:2012kn} for
many more $\pi N$ determinations).

The PWA of the Nijmegen group with the same $\chi$TPE
potential~\cite{Rentmeester:1999vw} but a different short distance
represention, included data up to $E_{\rm LAB} \le 500 {\rm MeV}$ and
gave $c_1=-4.4(3.4){\rm GeV}^{-1}$ which is different from our
findings that make it compatible with zero. We remind that our $NN$
analysis involves larger statistics (see Table \ref{chi2}) for $E_{\rm
  LAB} < 350 {\rm MeV}$ and hence the overall smaller uncertainties
are not surprising. Similarly to the Nijmegen
group~\cite{Rentmeester:1999vw}, we find a strong anti-correlation
between $c_1$ and $c_3$.  This allowed them to fix $c_1$ although the
error estimate is based on taking the $\pi N$ value for $c_1=-0.76(7)
{\rm GeV}^{-1}$. In our case, if we take $c_1=-0.76 {\rm GeV}^{-1}$ as
{\it input} we get after readjusting $c_3=-4.42(7){\rm GeV}^{-1}$ and
$c_4=4.47(16) {\rm GeV}^{-1}$ where, again, our errors are smaller
presumably due to larger statistics for $E_{\rm LAB} < 350 {\rm MeV}$.

The Nijmegen group found strong correlations of the chiral constants
with the pion-nucleon coupling
constant~\cite{Rentmeester:1999vw,Rentmeester:2003mf} when is
different from the recommended value
$f^2=0.075$~~\cite{Stoks:1992ja,deSwart:1997ep}.  This the fixed value
we took both in selection of data in our previous
work~\cite{Perez:2013mwa,Perez:2013jpa} as well as here.  We choose
not to change the coupling constant value as this will have some
impact on the data selection.

The recent values based on a $\chi$TPE fit up to $T_{\rm LAB} \le 125
{\rm MeV}$~\cite{Ekstrom:2013kea} are $2\sigma$ compatible with ours
although no errors are reported, so it is unclear how many of the
given digits are statistically significant.  We find that lowering the
energy range of the fit increases the uncertainties, making $\chi$TPE
statistically irrelevant in that energy range (see also the discussion
in Ref.~\cite{Amaro:2013zka} in connection to nuclear matrix
elements).  In Ref.~\cite{PavonValderrama:2005wv} an error analysis of
chiral constants from low energy NN data and the deuteron using the
N2LO $\chi$TPE based on a Monte Carlo, i.e. non-parametric, error
propagation, was carried out revealing a branching structure in the
three planes spanned by $c_1$, $c_3$ and $c_4$. It would be useful,
though computationally costly, to carry out such error analysis in our
scheme.

\begin{table}[t]
\caption{Summary of chiral constants $c_1,c_3 $and $c_4$ compared with determinations. The symbol $^*$ stands for input from $\pi N$.}
\centering
\label{tab:c1c3c4}       
	\begin{tabular}{c  c  c l  l  l}
\hline 
                            & Ref. & Source  & $c_1$      & $c_3$      & $c_4$      \\
\hline 
                            & &         & GeV$^{-1}$ & GeV$^{-1}$ & GeV$^{-1}$  \\
      \hline
      This Work             & & $NN$    &  -0.41(1.08)  & -4.66(60) & 4.31(17) \\
      Nijmegen              & \cite{Rentmeester:1999vw} & $pp$    & -0.76(07)$^*$ & -5.08(28) & 4.70(70) \\
      Nijmegen              & \cite{Rentmeester:2003mf} & $NN$    & -0.76(07)$^*$  & -4.88(10) & 3.92(22) \\
      E \& M a  & \cite{Entem:2003ft}& $NN$    & -0.81            & -3.40            & 3.40            \\
      E \& M b  & \cite{Entem:2003ft} & $NN$    & -0.81            & -3.20            & 5.40            \\
      PV \& RA     & \cite{PavonValderrama:2005wv} & $NN$    & -1.2(2)            & -2.6(1)            & 3.3(1)            \\
      Ekstr\"om et. al.     & \cite{Ekstrom:2013kea} & $NN$    & -0.92            & -3.89            & 4.31            \\
      B \& M & \cite{Buettiker:1999ap} & $\pi N$ & -0.81(15) & -4.69(1.34) & 3.40(4) \\
\hline 
	\end{tabular}
\end{table}

\begin{table}
\caption{$\chi^{2}$ values for different analyzes up to $E_{\rm LAB}
  \le 350 {\rm MeV}$, PWA93~\cite{Stoks:1993tb}, Nijm
  I~\cite{Stoks:1994wp}, Nijm II~\cite{Stoks:1994wp},
  Reid93~\cite{Stoks:1994wp} , AV18~\cite{Wiringa:1994wb},
  CD-Bonn~\cite{Machleidt:2000ge} , WJC1 and WJC2~\cite{Gross:2008ps},
  PWApp-TPE~\cite{Rentmeester:1999vw} and
  PWANN-TPE~\cite{Rentmeester:2003mf} ($^*$ means fit up to $E_{\rm
    LAB} \le 500 {\rm MeV}$), $\delta$-OPE~\cite{Perez:2013mwa} and
  $\delta$-TPE (present work).  $N_{pp}$ ($N_{np}$) denotes the number
  of $pp$ ($np$) data, $N_{\rm Dat}= N_{pp}+ N_{np} $ is the total
  number of fitted data, $N_{\rm Par}$ is the number of parameters and
  $\chi^2/\nu$ the corresponding $\chi^2$ per degree of freedom $\nu=
  N_{\rm Dat}-N_{\rm Par}$.}
\begin{tabular}{lccccccc}
Potential & $N_{pp}$ & $\chi^{2}_{pp}$ 
                                & $N_{np}$ & $\chi^{2}_{np}$ & $N_{\rm Dat} $ & $N_{\rm Par} $ &$\chi^2/\nu$   \\
\tableline
 PWA93   &   1787 &  1787  &  2526 & 2489 & 4313 & 39 &  1.01 \\
 NijmI   &   1787 &  1795  &  2526 & 2627 & 4313 & 41  & 1.03 \\
 NijmII   &  1787 &  1795  &  2526 & 2625 &  4313& 47  &1.03 \\
 Reid93   &  1787 & 1795   &  2526 & 2694   &   4313 & 50 & 1.03 \\
 Nijm93   &  1787 & 3175   &  2526 & 4848   &   4313 & 15 & 1.87 \\
 AV18   &  1787 &  1962  &  2526 & 2685 &   4313 & 40 & 1.09 \\
CDBonn   & 2932  & 2153  & 3058 & 3119 & 5990 & 43 &  1.02\\
WJC1   &  0 & -  & 3788 & 4015  & 3788  & 27 & 1.06 \\
WJC2   &  0 & -  & 3788 & 4015  & 3788  & 15  & 1.12 \\
pp$\chi$TPE$$  & 1951  & 1937  &  0 &  - & 1951 & 25 & 1.01 \\
NN$\chi$TPE$^*$  & 5109  & 5184  & 4786 & 4806 & 9895 & 73 & 1.02   \\
$\delta$-OPE   & 2996 &  3051 & 3717  &  3958 & 6711 & 46 & 1.05 \\
$\delta$-$\chi$TPE  & 2996  &  3177 & 3716 & 4058  & 6711 & 33 & 1.08 \\
\tableline
\end{tabular}
\label{chi2}
\end{table}

\section{Error propagation}
\label{sec:prop}

In table \ref{tab:DeuteronP} we show our results for the deuteron
static properties with their propagated errors and compared with our
previous PWA and other high quality potentials. 
As we see there is a trend to produce smaller errors in the $\chi$TPE
case as compared to the OPE result. The reason may be the slightly
larger $\chi^2$ value, which generically reduces the errors. The
compatibility with our previous OPE study is at the $2\sigma$-level.

\begin{figure*}[hpt]
\begin{center}
\epsfig{figure=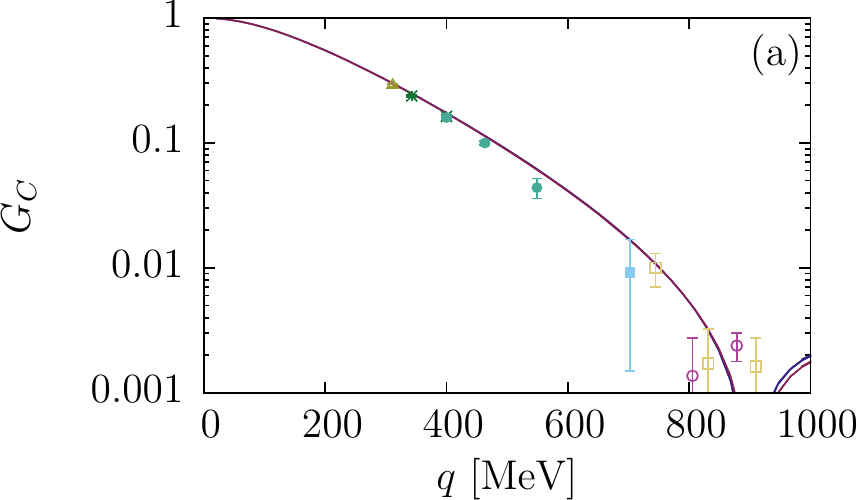,height=4cm,width=5cm}
\epsfig{figure=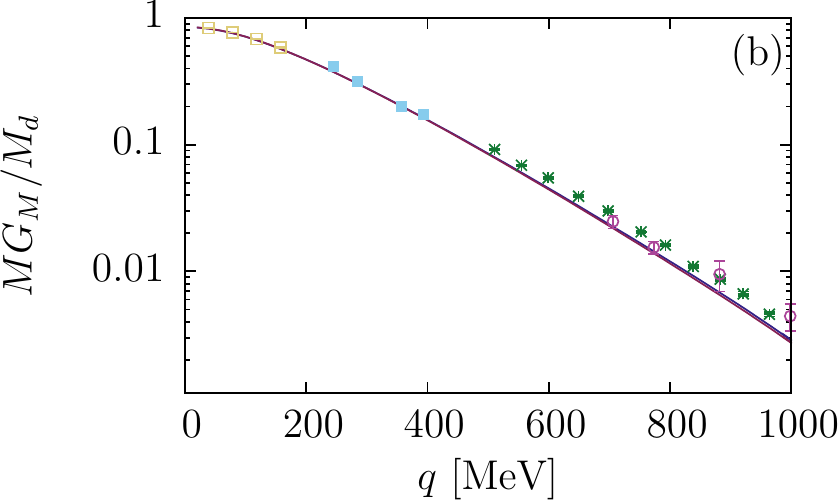,height=4cm,width=5cm}
\epsfig{figure=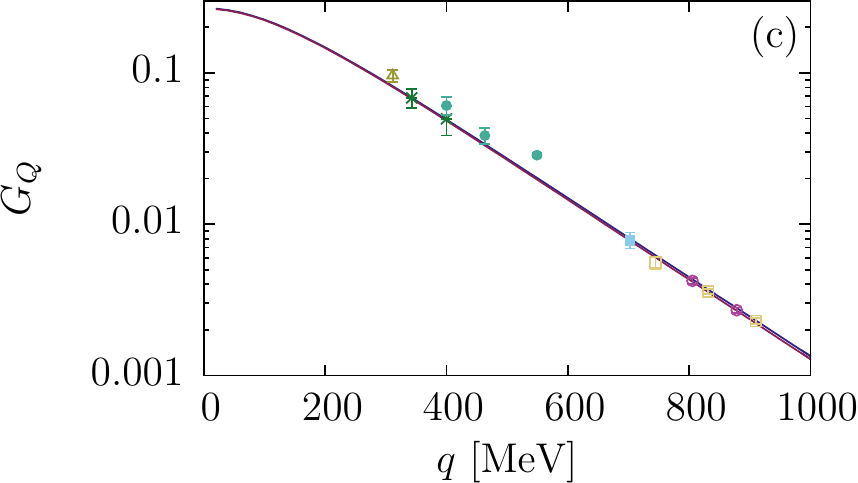,height=4cm,width=5cm}
\end{center}
\caption{Deuteron Form Factors with theoretical error bands obtained
  by propagating the uncertainties of the np+pp plus deuteron binding
  fit (see main text).  Note that the theoretical error is so tiny
  that the width of the bands cannot be seen at the scale of the
  figure.}
\label{FigFormFactors}
\end{figure*}

The Deuteron form factors $G_C(Q)$, $G_M(Q)$ and $G_Q(Q)$ (see
e.g.~\cite{Gilman:2001yh} for a review) are depicted in
Fig.~\ref{FigFormFactors} and come out with tiny error bands that
cannot be distinguished within the plot scale from the ones obtained
with  OPE only in our previous work~\cite{Perez:2013mwa}. 

\begin{figure*}[ttt]
\begin{center}
\epsfig{figure=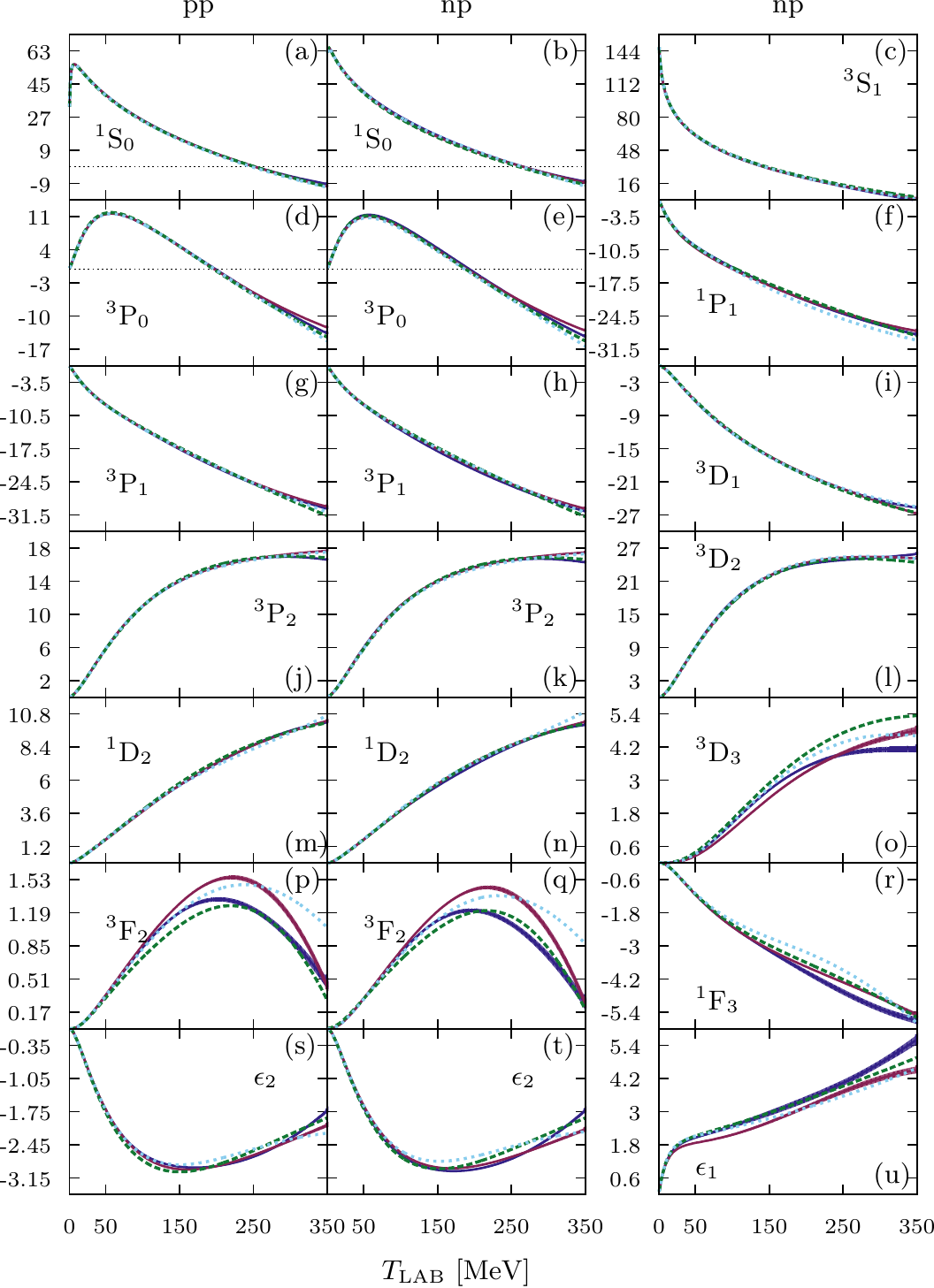,height=22cm,width=14.7cm}
\end{center}
\caption{Lowest np and pp phase shifts (in degrees) and their errors
  (solid band) from the present $\chi$TPE analysis (red) and our
  previous OPE (blue). We also compare with the Nijmegen 1993
  PWA~\cite{Stoks:1993tb} (light blue) and the
  AV18~\cite{Wiringa:1994wb} (green) as a function of the LAB energy
  (in MeV).}
\label{fig:all-phases}
\end{figure*}

In table \ref{tab:Fits} we show the strength operator coefficients
$V_{i,n}$ (see Eq.()) and their statistical uncertainties propagated
from the experimental data via the usual covariance matrix and
applying the linear transformation to the partial wave short distance
parameters$ \lambda_i$ discussed in Ref.~\cite{Perez:2013mwa}. With
these parameters and the covariance matrix it is possible to also
estimate and propagate statistical error bars for calculations made
with the $\delta$-shell potential.

\begin{figure*}[hpt]
\begin{center}
\epsfig{figure=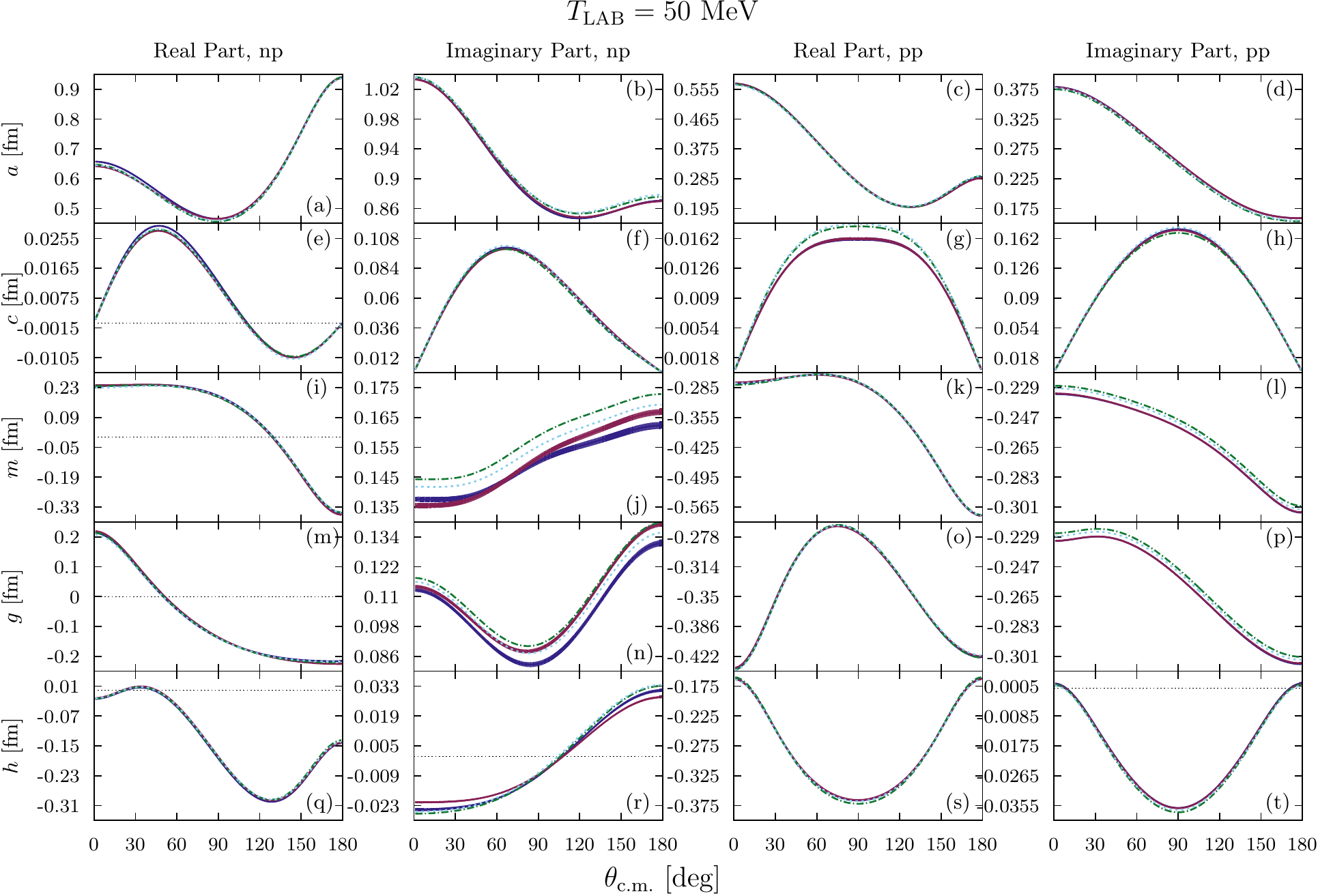,height=10cm,width=16cm}
\end{center}
\caption{Color on-line. np (left) and pp (right) Wolfenstein
  parameters (in fm) as a function of the CM angle (in degrees) and
  for $E_{\rm LAB}=50 {\rm MeV}$. We compare our fit (blue band) with
  the PWA~\cite{Stoks:1993tb} (dotted,magenta) and the AV18
  potential~\cite{Wiringa:1994wb} (dashed-dotted,black) which provided
  a $\chi^2/ {\rm d.o.f} \lesssim 1 $ for data before 1993.}
\label{FigWolfenstein050}
\end{figure*}

\begin{figure*}[hpt]
\begin{center}
\epsfig{figure=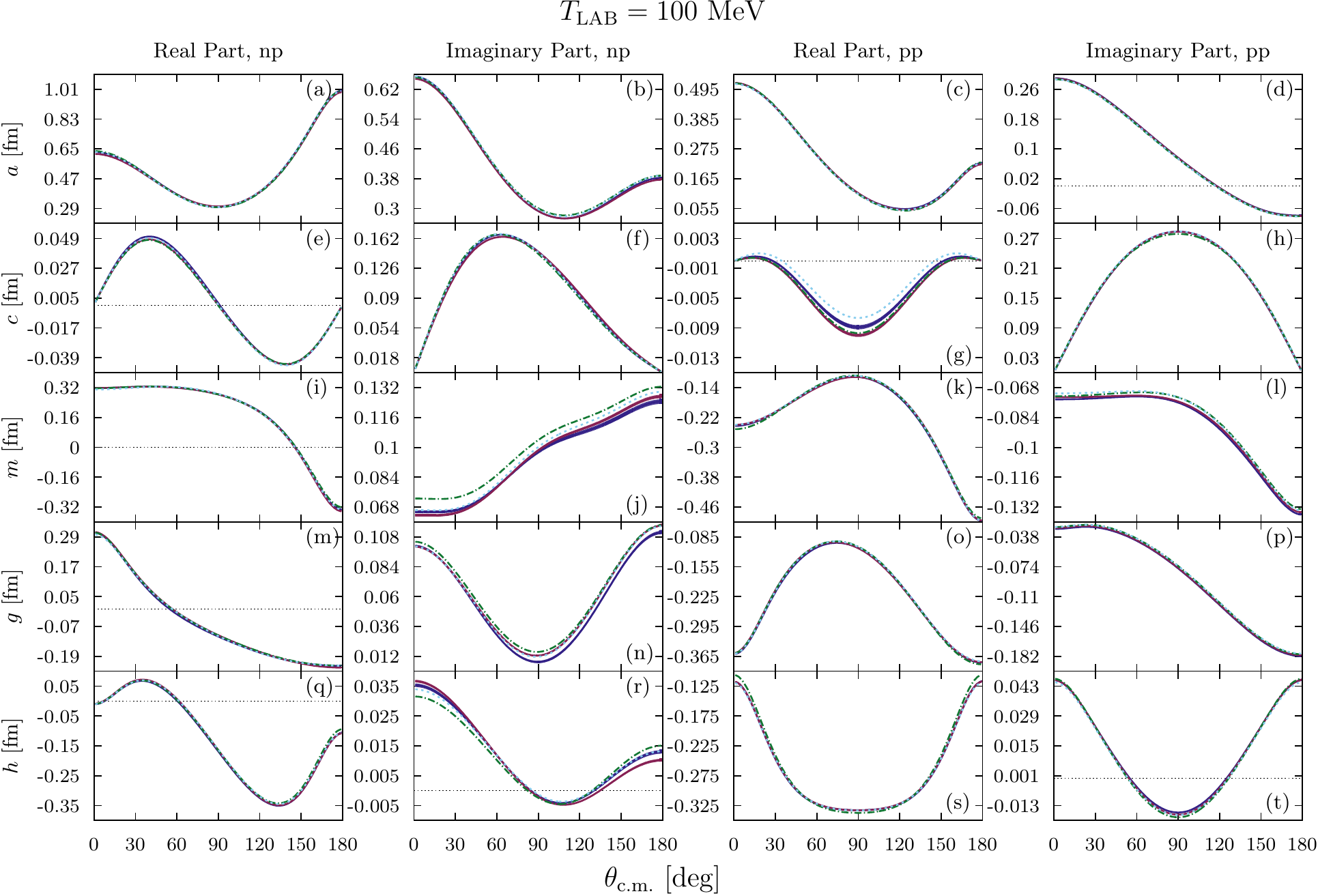,height=10cm,width=16cm}
\end{center}
\caption{Same as in Fig.~\ref{FigWolfenstein050} but for  $E_{\rm LAB}=100 {\rm
    MeV}$.}
\label{FigWolfenstein100}
\end{figure*}

\begin{figure*}[hpt]
\begin{center}
\epsfig{figure=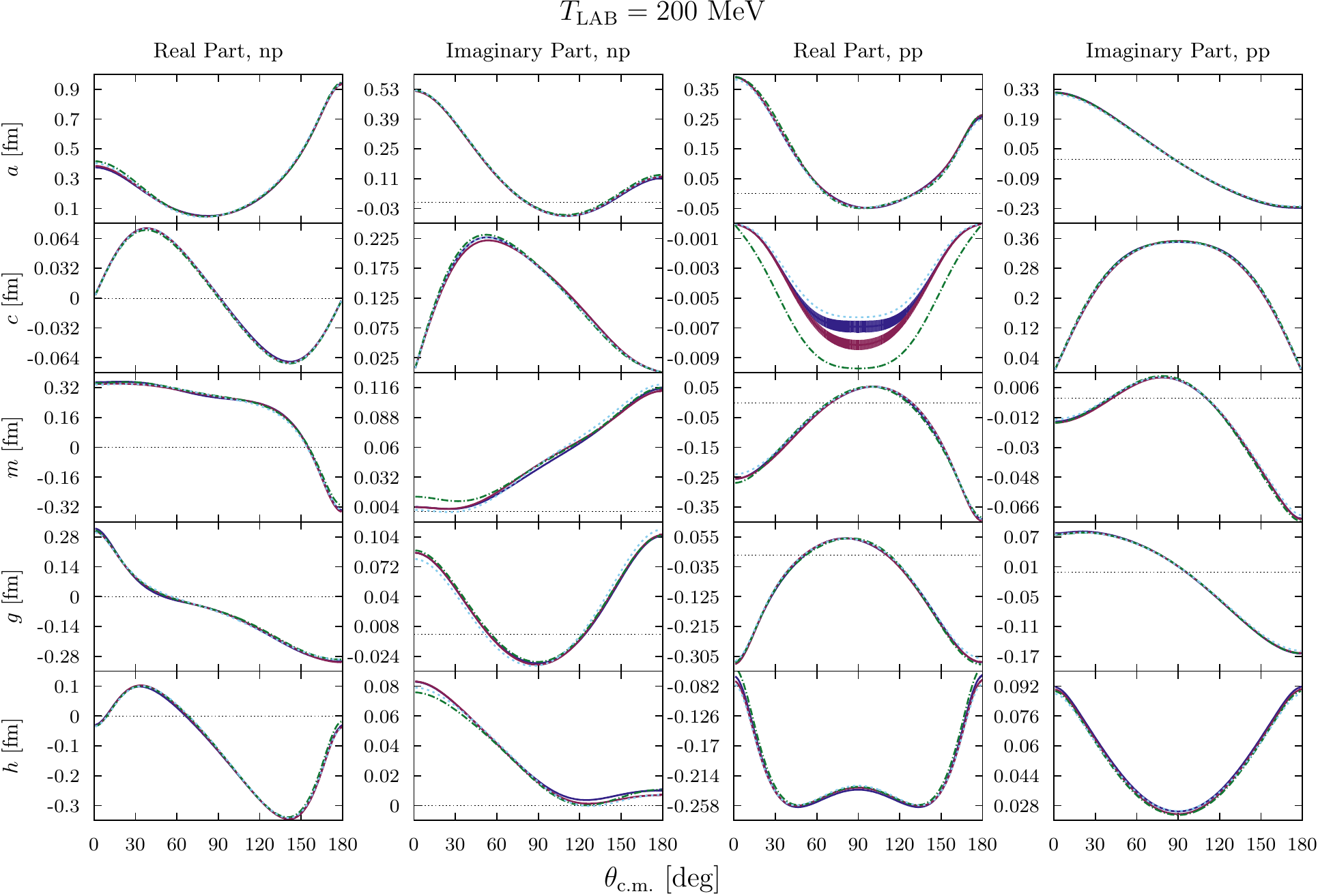,height=10cm,width=16cm}
\end{center}
\caption{Same as in Fig.~\ref{FigWolfenstein050} but for  $E_{\rm LAB}=200 {\rm
    MeV}$.}
\label{FigWolfenstein200}
\end{figure*}

\begin{figure*}[hpt]
\begin{center}
\epsfig{figure=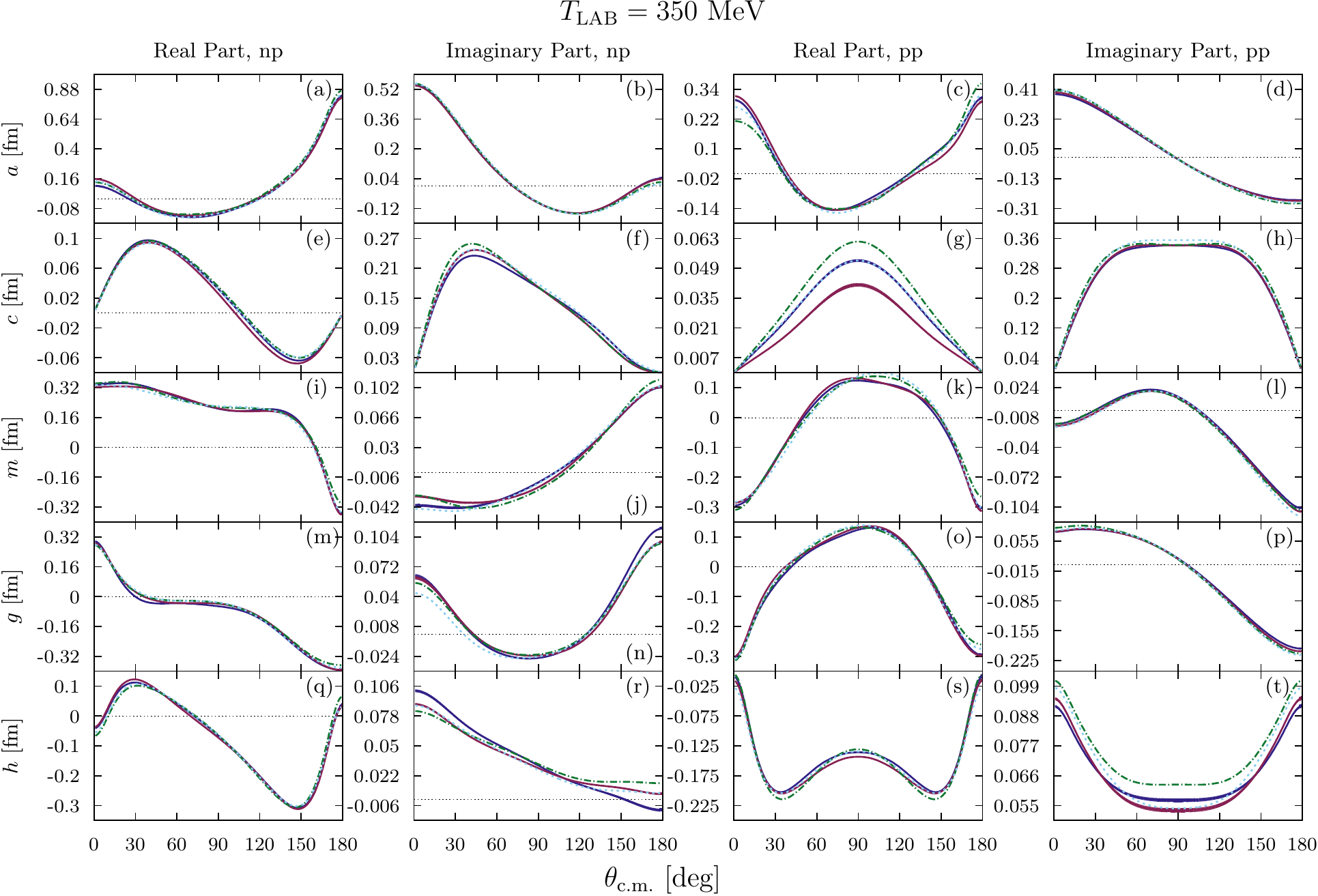,height=10cm,width=16cm}
\end{center}
\caption{Same as in Fig.~\ref{FigWolfenstein050} but for  $E_{\rm LAB}=350 {\rm
    MeV}$.}
\label{FigWolfenstein350}
\end{figure*}

In tables~\ref{tab:ppIsovectorPS}, \ref{tab:npIsovectorPS} and~,
\ref{tab:npIsoscalarPS} we show pp isovector, np isovector and np
isoscalar phaseshifts respectively with statistical errors extracted
from experimental data for the lowest partial waves at different
kinetic laboratory frame energy. A global overview can be appreciated
in Fig.~\ref{fig:all-phases} where we plot these phases.  For
comparison we also draw the phase shifts from our previous OPE
analysis~\cite{Perez:2013mwa,Perez:2013jpa}, the Nijmegen
PWA~\cite{Stoks:1993tb} and the AV18 potential~\cite{Wiringa:1994wb}.
As we see they agree within uncertainties for the lowest partial
waves.  Unfortunately the seminal Nijmegen group analysis of chiral
potentials~\cite{Rentmeester:1999vw,Rentmeester:2003mf}, did not
provide phases, so a direct comparison which would reflect the effect
of the different short distance parameterizations cannot be made. The
discrepancies apparent in higher partial waves among {\it all
  potentials} take also place in the scattering amplitude as shown in
Figs.~\ref{FigWolfenstein050},\ref{FigWolfenstein100},\ref{FigWolfenstein200},\ref{FigWolfenstein350}
and suggest the presence of some small systematic errors. The
systematic vs statistical errors dominance was already noted in
Refs.~\cite{NavarroPerez:2012vr,NavarroPerez:2012qf}.  A
non-parametric statistical analysis along the lines pursued in
Ref.~\cite{PavonValderrama:2005wv} for the complete database might
possibly shed light into this issue and is left for future research.

\section{Conclusion}
\label{sec:conc}

We summarize our points. The chiral constants $c_1$,$c_3$ and $c_4$
characterizing the $\chi$TPE potential at NNLO have been determined
with errors by analyzing NN scattering published data from 1950 till
2013 below $350 {\rm MeV}$ with a $\chi^2/\nu =1.08$. The values found
are in the bulk of other determinations, although our higher data
statistics allows to reduce previous error estimates based on NN
scattering data and the deuteron. At the same time we provide
quantitative error estimates of the short distance component of the
interaction hence allowing error propagation of the much used
$\chi$TPE interactions in Nuclear structure calculations.  We have
also provided extensive tables of phase-shifts with uncertainties
based on the present analysis. The verification and control of errors
in the NN interaction is an important test to check the validity and
statistical reliability of theoretical predictions with a prescribed
confidence level. Our results suggest that chiral interactions may
play an important role in Nuclear Structure calculations within the
errors inherited from the existing NN data.

\begin{table*}[ttt]
	\centering
	\caption{Deuteron static properties compared with empirical/recommended 
          values and high-quality potentials
          calculations. We list binding energy $E_d$, asymptotic D/S
          ratio $\eta$, asymptotic S-wave amplitude $A_S$, mean
          squared matter radius $r_m$, quadrupole moment $Q_D$, D-wave
          probability $P_D$ and inverse matter radius $\langle r^{-1}
          \rangle$.}
	\label{tab:DeuteronP}
	\begin{tabular*}{\textwidth}{@{\extracolsep{\fill}}l l l l l l l l l}
      \hline
      \hline
            & This work & Emp./Rec.\cite{VanDerLeun1982261,Borbély198517,Rodning:1990zz,Klarsfeld1986373,Bishop:1979zz,deSwart:1995ui} & $\delta$-shell~\cite{Perez:2013mwa} & Nijm I~\cite{Stoks:1994wp}   & Nijm II~\cite{Stoks:1994wp}  & Reid93~\cite{Stoks:1994wp}   & AV18~\cite{Wiringa:1994wb} & CD-Bonn~\cite{Machleidt:2000ge}  \\
      \hline
		$E_d$(MeV)              & Input       & 2.224575(9)    & Input       & Input    & Input    & Input    & Input  & Input  \\
		$\eta$                  & 0.02473(4)  & 0.0256(5)      & 0.02493(8)  & 0.02534  & 0.02521  & 0.02514  & 0.0250 & 0.0256 \\
		$A_S ({\rm fm}^{1/2})$  & 0.8854(2)   & 0.8845(8)     & 0.8829(4)   & 0.8841   & 0.8845   & 0.8853   & 0.8850 & 0.8846 \\
		$r_m ({\rm fm})$        & 1.9689(4)   & 1.971(6)       & 1.9645(9)   & 1.9666   & 1.9675   & 1.9686   & 1.967  &  1.966 \\
		$Q_D ({\rm fm}^{2}) $   & 0.2658(5)   & 0.2859(3)      & 0.2679(9)   & 0.2719   & 0.2707   & 0.2703   & 0.270  & 0.270  \\
		$P_D$                   & 5.30(3)     & 5.67(4)        & 5.62(5)     & 5.664    & 5.635    & 5.699    & 5.76   & 4.85   \\
		$\langle r^{-1} \rangle ({\rm fm}^{-1})$& 0.4542(2) &  & 0.4540(5)   &          & 0.4502   & 0.4515   &        &        \\
      \hline \hline
	\end{tabular*}
\end{table*}

\begingroup
\begin{table}[htb]
 \caption{\label{tab:Fits} Delta-shell potential parameters $V_{i,n}
        $ (in ${\rm fm}^{-1}$) with their errors for all operators.
        We take $N=3$ equidistant points with $\Delta r
        = 0.6$fm. Rows marked with $^*$ indicates that the corresponding strengths 
        coefficients are not independent. In the first line we provide the
        central component of the delta shells corresponding to the EM
        effects below $r_c = 1.8 {\rm fm}$. These parameters remain
        fixed within the fitting process.}
  \begin{ruledtabular}      
	\begin{tabular*}{\columnwidth}{@{\extracolsep{\fill}}c *{2}{D{.}{.}{2.7}} *{1}{D{.}{.}{2.9}} }
 Operator  & \multicolumn{1}{c}{$V_{1,x}$} & \multicolumn{1}{c}{$V_{2,x}$} & \multicolumn{1}{c}{$V_{3,x}$} \\
       & \multicolumn{1}{c}{$r_1=0.6$fm} & \multicolumn{1}{c}{$r_2=1.2$fm} & \multicolumn{1}{c}{$r_3=1.8$fm} \\
            \hline
$V_C[{\rm pp}]_{\rm EM}$  & 0.0072555  &   0.0065592  &   0.0016129  \\ 
\noalign{\smallskip}
 $c$               &      0.395(2)   &   -0.022(3)    &  -0.0119(9)  \\
 $\tau$            &      0.030(2)   &   -0.036(1)    &   0.0025(2)  \\
 $\sigma$          &     -0.021(2)   &    0.041(1)    &  -0.0002(2)  \\
 $\sigma\tau$      &     -0.0410(8)  &    0.0417(7)   &   0.0021(1)  \\
 $t$               &      0.0        &   -0.002(2)    &   0.0008(2)  \\
 $t \tau$          &      0.0        &    0.1029(7)   &   0.0043(1)  \\
 $ls$              &     -0.1253(5)  &   -0.117(3)    &   0.0003(3)  \\
 $ls \tau$         &     -0.0418(2)  &   -0.025(1)    &  -0.0016(1)  \\
 $l2$              &     -0.2416(5)  &    0.022(3)    &   0.0005(2)  \\
 $l2 \tau$         &     -0.0636(3)  &   -0.008(1)    &   0.00012(8) \\
 $l2 \sigma$       &     -0.0551(4)  &    0.000(1)    &   0.00003(9) \\
 $l2 \sigma\tau$   &     -0.0127(1)  &   -0.0027(4)   &  -0.00001(3) \\
 $ls2$             &      0.1614(4)  &    0.030(5)    &  -0.0012(4)  \\
 $ls2 \tau$        &      0.0538(1)  &    0.024(2)    &  -0.0010(1)  \\
 $T$               &      0.006(1)   &   -0.0003(2)   &  -0.00005(9) \\
 $\sigma T^*$      &     -0.006(1)   &    0.0003(2)   &   0.00005(9) \\
 $t T^*$           &      0.0        &    0.0         &   0.0        \\
 $\tau z^*$        &      0.0        &    0.0         &   0.0        \\
 $\sigma\tau z^*$  &      0.0        &    0.0         &   0.0        \\
 $l2 T^*$          &     -0.0010(2)  &    0.00004(4)  &   0.00001(1) \\
 $l2 \sigma T^*$   &      0.0010(2)  &   -0.00004(4)  &  -0.00001(1) \\
	\end{tabular*}
 \end{ruledtabular}
\end{table}
\endgroup
     
\begin{table*}
 \footnotesize
 \caption{\label{tab:ppIsovectorPS} pp isovector phaseshifts.}
 \begin{ruledtabular}
 \begin{tabular*}{\textwidth}{@{\extracolsep{\fill}} r *{12}{D{.}{.}{3.3}}}
$E_{\rm LAB}$&\multicolumn{1}{c}{$^1S_0$}&\multicolumn{1}{c}{$^1D_2$}&\multicolumn{1}{c}{$^1G_4$}&\multicolumn{1}{c}{$^3P_0$}&\multicolumn{1}{c}{$^3P_1$}&\multicolumn{1}{c}{$^3F_3$}&\multicolumn{1}{c}{$^3P_2$}&\multicolumn{1}{c}{$\epsilon_2$}&\multicolumn{1}{c}{$^3F_2$}&\multicolumn{1}{c}{$^3F_4$}&\multicolumn{1}{c}{$\epsilon_4$}&\multicolumn{1}{c}{$^3H_4$}\\ 
  \hline 
  1 &    32.654 &     0.001 &     0.000 &     0.133 &    -0.079 &    -0.000 &     0.014 &    -0.001 &     0.000 &     0.000 &    -0.000 &     0.000\\
    & \pm 0.003 & \pm 0.000 & \pm 0.000 & \pm 0.000 & \pm 0.000 & \pm 0.000 & \pm 0.000 & \pm 0.000 & \pm 0.000 & \pm 0.000 & \pm 0.000 & \pm 0.000\\
  5 &    54.879 &     0.042 &     0.000 &     1.578 &    -0.886 &    -0.004 &     0.216 &    -0.051 &     0.002 &     0.000 &    -0.000 &     0.000\\
    & \pm 0.005 & \pm 0.000 & \pm 0.000 & \pm 0.002 & \pm 0.001 & \pm 0.000 & \pm 0.001 & \pm 0.000 & \pm 0.000 & \pm 0.000 & \pm 0.000 & \pm 0.000\\
 10 &    55.313 &     0.165 &     0.003 &     3.726 &    -2.024 &    -0.031 &     0.652 &    -0.200 &     0.013 &     0.001 &    -0.003 &     0.000\\
    & \pm 0.007 & \pm 0.000 & \pm 0.000 & \pm 0.005 & \pm 0.002 & \pm 0.000 & \pm 0.002 & \pm 0.000 & \pm 0.000 & \pm 0.000 & \pm 0.000 & \pm 0.000\\
 25 &    48.852 &     0.690 &     0.040 &     8.590 &    -4.837 &    -0.230 &     2.487 &    -0.806 &     0.106 &     0.021 &    -0.049 &     0.004\\
    & \pm 0.010 & \pm 0.001 & \pm 0.000 & \pm 0.016 & \pm 0.005 & \pm 0.000 & \pm 0.005 & \pm 0.000 & \pm 0.000 & \pm 0.000 & \pm 0.000 & \pm 0.000\\
 50 &    39.176 &     1.679 &     0.154 &    11.532 &    -8.176 &    -0.690 &     5.840 &    -1.704 &     0.346 &     0.112 &    -0.196 &     0.026\\
    & \pm 0.015 & \pm 0.003 & \pm 0.000 & \pm 0.029 & \pm 0.009 & \pm 0.001 & \pm 0.008 & \pm 0.001 & \pm 0.001 & \pm 0.001 & \pm 0.000 & \pm 0.000\\
100 &    25.305 &     3.727 &     0.428 &     9.473 &   -13.161 &    -1.527 &    11.026 &    -2.684 &     0.857 &     0.478 &    -0.546 &     0.111\\
    & \pm 0.026 & \pm 0.007 & \pm 0.001 & \pm 0.041 & \pm 0.016 & \pm 0.005 & \pm 0.014 & \pm 0.005 & \pm 0.004 & \pm 0.003 & \pm 0.000 & \pm 0.000\\
150 &    15.179 &     5.636 &     0.709 &     4.665 &   -17.426 &    -2.143 &    14.037 &    -2.967 &     1.282 &     1.007 &    -0.863 &     0.222\\
    & \pm 0.038 & \pm 0.012 & \pm 0.002 & \pm 0.045 & \pm 0.022 & \pm 0.014 & \pm 0.018 & \pm 0.008 & \pm 0.009 & \pm 0.007 & \pm 0.000 & \pm 0.001\\
200 &     7.083 &     7.239 &     0.998 &    -0.384 &   -21.266 &    -2.582 &    15.719 &    -2.902 &     1.526 &     1.604 &    -1.128 &     0.347\\
    & \pm 0.053 & \pm 0.016 & \pm 0.004 & \pm 0.050 & \pm 0.032 & \pm 0.026 & \pm 0.021 & \pm 0.013 & \pm 0.017 & \pm 0.011 & \pm 0.001 & \pm 0.002\\
250 &     0.292 &     8.512 &     1.288 &    -5.029 &   -24.654 &    -2.845 &    16.696 &    -2.669 &     1.500 &     2.187 &    -1.345 &     0.483\\
    & \pm 0.071 & \pm 0.020 & \pm 0.008 & \pm 0.061 & \pm 0.048 & \pm 0.039 & \pm 0.028 & \pm 0.019 & \pm 0.026 & \pm 0.016 & \pm 0.002 & \pm 0.003\\
300 &    -5.563 &     9.504 &     1.567 &    -9.068 &   -27.515 &    -2.800 &    17.298 &    -2.363 &     1.144 &     2.689 &    -1.519 &     0.632\\
    & \pm 0.094 & \pm 0.030 & \pm 0.014 & \pm 0.078 & \pm 0.071 & \pm 0.046 & \pm 0.039 & \pm 0.028 & \pm 0.036 & \pm 0.024 & \pm 0.003 & \pm 0.005\\
350 &   -10.689 &    10.286 &     1.818 &   -12.438 &   -29.766 &    -1.907 &    17.682 &    -2.040 &     0.430 &     3.065 &    -1.653 &     0.794\\
    & \pm 0.120 & \pm 0.050 & \pm 0.023 & \pm 0.098 & \pm 0.098 & \pm 0.085 & \pm 0.050 & \pm 0.038 & \pm 0.045 & \pm 0.042 & \pm 0.004 & \pm 0.008\\
 \end{tabular*}
 \end{ruledtabular}
\end{table*}

\begin{table*}
 \footnotesize
 \caption{\label{tab:npIsovectorPS}np isovector phaseshifts.}
 \begin{ruledtabular}
  \begin{tabular*}{\textwidth}{@{\extracolsep{\fill}} r *{12}{D{.}{.}{3.3}}}
$E_{\rm LAB}$&\multicolumn{1}{c}{$^1S_0$}&\multicolumn{1}{c}{$^1D_2$}&\multicolumn{1}{c}{$^1G_4$}&\multicolumn{1}{c}{$^3P_0$}&\multicolumn{1}{c}{$^3P_1$}&\multicolumn{1}{c}{$^3F_3$}&\multicolumn{1}{c}{$^3P_2$}&\multicolumn{1}{c}{$\epsilon_2$}&\multicolumn{1}{c}{$^3F_2$}&\multicolumn{1}{c}{$^3F_4$}&\multicolumn{1}{c}{$\epsilon_4$}&\multicolumn{1}{c}{$^3H_4$}\\ 
  \hline 
  1 &    62.083 &     0.001 &     0.000 &     0.177 &    -0.106 &    -0.000 &     0.022 &    -0.001 &     0.000 &     0.000 &    -0.000 &     0.000\\
    & \pm 0.015 & \pm 0.000 & \pm 0.000 & \pm 0.000 & \pm 0.000 & \pm 0.000 & \pm 0.000 & \pm 0.000 & \pm 0.000 & \pm 0.000 & \pm 0.000 & \pm 0.000\\
  5 &    63.652 &     0.041 &     0.000 &     1.617 &    -0.918 &    -0.004 &     0.255 &    -0.048 &     0.002 &     0.000 &    -0.000 &     0.000\\
    & \pm 0.038 & \pm 0.000 & \pm 0.000 & \pm 0.002 & \pm 0.001 & \pm 0.000 & \pm 0.001 & \pm 0.000 & \pm 0.000 & \pm 0.000 & \pm 0.000 & \pm 0.000\\
 10 &    59.983 &     0.156 &     0.002 &     3.649 &    -2.021 &    -0.026 &     0.719 &    -0.182 &     0.011 &     0.001 &    -0.003 &     0.000\\
    & \pm 0.056 & \pm 0.000 & \pm 0.000 & \pm 0.006 & \pm 0.002 & \pm 0.000 & \pm 0.002 & \pm 0.000 & \pm 0.000 & \pm 0.000 & \pm 0.000 & \pm 0.000\\
 25 &    50.895 &     0.672 &     0.033 &     8.197 &    -4.777 &    -0.198 &     2.590 &    -0.748 &     0.091 &     0.018 &    -0.039 &     0.003\\
    & \pm 0.091 & \pm 0.001 & \pm 0.000 & \pm 0.016 & \pm 0.005 & \pm 0.000 & \pm 0.005 & \pm 0.000 & \pm 0.000 & \pm 0.000 & \pm 0.000 & \pm 0.000\\
 50 &    40.410 &     1.683 &     0.136 &    10.904 &    -8.122 &    -0.617 &     5.948 &    -1.619 &     0.310 &     0.101 &    -0.168 &     0.021\\
    & \pm 0.126 & \pm 0.003 & \pm 0.000 & \pm 0.029 & \pm 0.009 & \pm 0.001 & \pm 0.008 & \pm 0.001 & \pm 0.001 & \pm 0.001 & \pm 0.000 & \pm 0.000\\
100 &    26.350 &     3.787 &     0.402 &     8.750 &   -13.211 &    -1.410 &    11.076 &    -2.611 &     0.797 &     0.452 &    -0.495 &     0.095\\
    & \pm 0.160 & \pm 0.007 & \pm 0.001 & \pm 0.041 & \pm 0.016 & \pm 0.005 & \pm 0.014 & \pm 0.005 & \pm 0.004 & \pm 0.003 & \pm 0.000 & \pm 0.000\\
150 &    16.345 &     5.717 &     0.690 &     3.945 &   -17.588 &    -2.008 &    14.011 &    -2.930 &     1.204 &     0.970 &    -0.803 &     0.196\\
    & \pm 0.178 & \pm 0.013 & \pm 0.003 & \pm 0.046 & \pm 0.023 & \pm 0.014 & \pm 0.018 & \pm 0.009 & \pm 0.009 & \pm 0.007 & \pm 0.000 & \pm 0.001\\
200 &     8.437 &     7.306 &     0.991 &    -1.088 &   -21.525 &    -2.436 &    15.630 &    -2.904 &     1.430 &     1.559 &    -1.066 &     0.314\\
    & \pm 0.205 & \pm 0.017 & \pm 0.009 & \pm 0.051 & \pm 0.032 & \pm 0.027 & \pm 0.021 & \pm 0.013 & \pm 0.017 & \pm 0.011 & \pm 0.001 & \pm 0.002\\
250 &     1.858 &     8.541 &     1.298 &    -5.718 &   -24.988 &    -2.682 &    16.558 &    -2.707 &     1.384 &     2.132 &    -1.284 &     0.445\\
    & \pm 0.255 & \pm 0.020 & \pm 0.020 & \pm 0.062 & \pm 0.049 & \pm 0.039 & \pm 0.029 & \pm 0.020 & \pm 0.027 & \pm 0.016 & \pm 0.002 & \pm 0.003\\
300 &    -3.772 &     9.485 &     1.596 &    -9.742 &   -27.903 &    -2.581 &    17.125 &    -2.431 &     1.003 &     2.621 &    -1.461 &     0.589\\
    & \pm 0.328 & \pm 0.031 & \pm 0.037 & \pm 0.079 & \pm 0.072 & \pm 0.047 & \pm 0.040 & \pm 0.028 & \pm 0.036 & \pm 0.025 & \pm 0.003 & \pm 0.005\\
350 &    -8.658 &    10.218 &     1.868 &   -13.098 &   -30.187 &    -1.432 &    17.482 &    -2.130 &     0.261 &     2.982 &    -1.600 &     0.749\\
    & \pm 0.418 & \pm 0.052 & \pm 0.060 & \pm 0.100 & \pm 0.100 & \pm 0.114 & \pm 0.051 & \pm 0.039 & \pm 0.045 & \pm 0.043 & \pm 0.004 & \pm 0.008\\
 \end{tabular*}
 \end{ruledtabular}
\end{table*}

\begin{table*}
 \footnotesize
 \caption{\label{tab:npIsoscalarPS}np isoscalar phaseshifts.}
 \begin{ruledtabular}
 \begin{tabular*}{\textwidth}{@{\extracolsep{\fill}} r *{12}{D{.}{.}{3.3}}}
 $E_{\rm LAB}$&\multicolumn{1}{c}{$^1P_1$}&\multicolumn{1}{c}{$^1F_3$}&\multicolumn{1}{c}{$^3D_2$}&\multicolumn{1}{c}{$^3G_4$}&\multicolumn{1}{c}{$^3S_1$}&\multicolumn{1}{c}{$\epsilon_1$}&\multicolumn{1}{c}{$^3D_1$}&\multicolumn{1}{c}{$^3D_3$}&\multicolumn{1}{c}{$\epsilon_3$}&\multicolumn{1}{c}{$^3G_3$}\\ 
  \hline 
  1 &    -0.189 &    -0.000 &     0.006 &     0.000 &   147.716 &     0.103 &    -0.005 &     0.000 &     0.000 &    -0.000\\
    & \pm 0.000 & \pm 0.000 & \pm 0.000 & \pm 0.000 & \pm 0.007 & \pm 0.000 & \pm 0.000 & \pm 0.000 & \pm 0.000 & \pm 0.000\\
  5 &    -1.528 &    -0.010 &     0.217 &     0.001 &   118.103 &     0.641 &    -0.178 &     0.002 &     0.012 &    -0.000\\
    & \pm 0.002 & \pm 0.000 & \pm 0.000 & \pm 0.000 & \pm 0.015 & \pm 0.002 & \pm 0.000 & \pm 0.000 & \pm 0.000 & \pm 0.000\\
 10 &    -3.148 &    -0.063 &     0.842 &     0.012 &   102.497 &     1.088 &    -0.665 &     0.004 &     0.079 &    -0.003\\
    & \pm 0.006 & \pm 0.000 & \pm 0.000 & \pm 0.000 & \pm 0.022 & \pm 0.005 & \pm 0.001 & \pm 0.000 & \pm 0.000 & \pm 0.000\\
 25 &    -6.607 &    -0.421 &     3.689 &     0.170 &    80.418 &     1.634 &    -2.756 &     0.032 &     0.553 &    -0.053\\
    & \pm 0.018 & \pm 0.000 & \pm 0.003 & \pm 0.000 & \pm 0.034 & \pm 0.013 & \pm 0.004 & \pm 0.001 & \pm 0.000 & \pm 0.000\\
 50 &   -10.086 &    -1.147 &     8.905 &     0.724 &    62.438 &     1.865 &    -6.344 &     0.254 &     1.613 &    -0.263\\
    & \pm 0.037 & \pm 0.001 & \pm 0.012 & \pm 0.000 & \pm 0.045 & \pm 0.024 & \pm 0.011 & \pm 0.006 & \pm 0.000 & \pm 0.000\\
100 &   -14.557 &    -2.310 &    17.147 &     2.200 &    42.803 &     2.161 &   -12.088 &     1.219 &     3.502 &    -0.971\\
    & \pm 0.067 & \pm 0.006 & \pm 0.039 & \pm 0.001 & \pm 0.053 & \pm 0.040 & \pm 0.024 & \pm 0.020 & \pm 0.003 & \pm 0.001\\
150 &   -18.070 &    -3.123 &    21.986 &     3.724 &    30.365 &     2.614 &   -16.318 &     2.348 &     4.842 &    -1.849\\
    & \pm 0.092 & \pm 0.013 & \pm 0.060 & \pm 0.004 & \pm 0.053 & \pm 0.053 & \pm 0.034 & \pm 0.035 & \pm 0.008 & \pm 0.003\\
200 &   -21.136 &    -3.756 &    24.371 &     5.166 &    21.054 &     3.166 &   -19.610 &     3.297 &     5.734 &    -2.765\\
    & \pm 0.121 & \pm 0.022 & \pm 0.071 & \pm 0.011 & \pm 0.055 & \pm 0.068 & \pm 0.039 & \pm 0.049 & \pm 0.016 & \pm 0.008\\
250 &   -23.794 &    -4.322 &    25.283 &     6.483 &    13.508 &     3.722 &   -22.324 &     3.988 &     6.304 &    -3.653\\
    & \pm 0.153 & \pm 0.036 & \pm 0.086 & \pm 0.025 & \pm 0.067 & \pm 0.085 & \pm 0.045 & \pm 0.067 & \pm 0.027 & \pm 0.017\\
300 &   -25.998 &    -4.891 &    25.404 &     7.644 &     7.100 &     4.196 &   -24.680 &     4.465 &     6.652 &    -4.485\\
    & \pm 0.186 & \pm 0.055 & \pm 0.116 & \pm 0.045 & \pm 0.091 & \pm 0.105 & \pm 0.063 & \pm 0.095 & \pm 0.039 & \pm 0.029\\
350 &   -27.677 &    -5.510 &    25.161 &     8.615 &     1.501 &     4.511 &   -26.802 &     4.808 &     6.852 &    -5.251\\
    & \pm 0.218 & \pm 0.080 & \pm 0.159 & \pm 0.072 & \pm 0.121 & \pm 0.124 & \pm 0.096 & \pm 0.131 & \pm 0.053 & \pm 0.045\\
 \end{tabular*}
 \end{ruledtabular}
\end{table*}

\begin{thebibliography}{37}
\expandafter\ifx\csname natexlab\endcsname\relax\def\natexlab#1{#1}\fi
\expandafter\ifx\csname bibnamefont\endcsname\relax
  \def\bibnamefont#1{#1}\fi
\expandafter\ifx\csname bibfnamefont\endcsname\relax
  \def\bibfnamefont#1{#1}\fi
\expandafter\ifx\csname citenamefont\endcsname\relax
  \def\citenamefont#1{#1}\fi
\expandafter\ifx\csname url\endcsname\relax
  \def\url#1{\texttt{#1}}\fi
\expandafter\ifx\csname urlprefix\endcsname\relax\def\urlprefix{URL }\fi
\providecommand{\bibinfo}[2]{#2}
\providecommand{\eprint}[2][]{\url{#2}}

\bibitem[{\citenamefont{Weinberg}(1990)}]{Weinberg:1990rz}
\bibinfo{author}{\bibfnamefont{S.}~\bibnamefont{Weinberg}},
  \bibinfo{journal}{Phys.Lett.} \textbf{\bibinfo{volume}{B251}},
  \bibinfo{pages}{288} (\bibinfo{year}{1990}).

\bibitem[{\citenamefont{Ordonez and van Kolck}(1992)}]{Ordonez:1992xp}
\bibinfo{author}{\bibfnamefont{C.}~\bibnamefont{Ordonez}} \bibnamefont{and}
  \bibinfo{author}{\bibfnamefont{U.}~\bibnamefont{van Kolck}},
  \bibinfo{journal}{Phys.Lett.} \textbf{\bibinfo{volume}{B291}},
  \bibinfo{pages}{459} (\bibinfo{year}{1992}).

\bibitem[{\citenamefont{Ordonez et~al.}(1996)\citenamefont{Ordonez, Ray, and
  van Kolck}}]{Ordonez:1995rz}
\bibinfo{author}{\bibfnamefont{C.}~\bibnamefont{Ordonez}},
  \bibinfo{author}{\bibfnamefont{L.}~\bibnamefont{Ray}}, \bibnamefont{and}
  \bibinfo{author}{\bibfnamefont{U.}~\bibnamefont{van Kolck}},
  \bibinfo{journal}{Phys.Rev.} \textbf{\bibinfo{volume}{C53}},
  \bibinfo{pages}{2086} (\bibinfo{year}{1996}), \eprint{hep-ph/9511380}.

\bibitem[{\citenamefont{Epelbaum et~al.}(2009)\citenamefont{Epelbaum, Hammer,
  and Meissner}}]{Epelbaum:2008ga}
\bibinfo{author}{\bibfnamefont{E.}~\bibnamefont{Epelbaum}},
  \bibinfo{author}{\bibfnamefont{H.-W.} \bibnamefont{Hammer}},
  \bibnamefont{and} \bibinfo{author}{\bibfnamefont{U.-G.}
  \bibnamefont{Meissner}}, \bibinfo{journal}{Rev.Mod.Phys.}
  \textbf{\bibinfo{volume}{81}}, \bibinfo{pages}{1773} (\bibinfo{year}{2009}),
  \eprint{0811.1338}.

\bibitem[{\citenamefont{Machleidt and Entem}(2011)}]{Machleidt:2011zz}
\bibinfo{author}{\bibfnamefont{R.}~\bibnamefont{Machleidt}} \bibnamefont{and}
  \bibinfo{author}{\bibfnamefont{D.}~\bibnamefont{Entem}},
  \bibinfo{journal}{Phys.Rept.} \textbf{\bibinfo{volume}{503}},
  \bibinfo{pages}{1} (\bibinfo{year}{2011}), \eprint{1105.2919}.

\bibitem[{\citenamefont{Kaiser et~al.}(1997)\citenamefont{Kaiser, Brockmann,
  and Weise}}]{Kaiser:1997mw}
\bibinfo{author}{\bibfnamefont{N.}~\bibnamefont{Kaiser}},
  \bibinfo{author}{\bibfnamefont{R.}~\bibnamefont{Brockmann}},
  \bibnamefont{and} \bibinfo{author}{\bibfnamefont{W.}~\bibnamefont{Weise}},
  \bibinfo{journal}{Nucl.Phys.} \textbf{\bibinfo{volume}{A625}},
  \bibinfo{pages}{758} (\bibinfo{year}{1997}), \eprint{nucl-th/9706045}.

\bibitem[{\citenamefont{Epelbaum}(2006)}]{Epelbaum:2005pn}
\bibinfo{author}{\bibfnamefont{E.}~\bibnamefont{Epelbaum}},
  \bibinfo{journal}{Prog.Part.Nucl.Phys.} \textbf{\bibinfo{volume}{57}},
  \bibinfo{pages}{654} (\bibinfo{year}{2006}), \eprint{nucl-th/0509032}.

\bibitem[{\citenamefont{Stoks et~al.}(1993)\citenamefont{Stoks, Kompl,
  Rentmeester, and de~Swart}}]{Stoks:1993tb}
\bibinfo{author}{\bibfnamefont{V.~G.~J.} \bibnamefont{Stoks}},
  \bibinfo{author}{\bibfnamefont{R.~A.~M.} \bibnamefont{Kompl}},
  \bibinfo{author}{\bibfnamefont{M.~C.~M.} \bibnamefont{Rentmeester}},
  \bibnamefont{and} \bibinfo{author}{\bibfnamefont{J.~J.}
  \bibnamefont{de~Swart}}, \bibinfo{journal}{Phys. Rev.}
  \textbf{\bibinfo{volume}{C48}}, \bibinfo{pages}{792} (\bibinfo{year}{1993}).

\bibitem[{\citenamefont{Stoks et~al.}(1994)\citenamefont{Stoks, Klomp,
  Terheggen, and de~Swart}}]{Stoks:1994wp}
\bibinfo{author}{\bibfnamefont{V.~G.~J.} \bibnamefont{Stoks}},
  \bibinfo{author}{\bibfnamefont{R.~A.~M.} \bibnamefont{Klomp}},
  \bibinfo{author}{\bibfnamefont{C.~P.~F.} \bibnamefont{Terheggen}},
  \bibnamefont{and} \bibinfo{author}{\bibfnamefont{J.~J.}
  \bibnamefont{de~Swart}}, \bibinfo{journal}{Phys. Rev.}
  \textbf{\bibinfo{volume}{C49}}, \bibinfo{pages}{2950} (\bibinfo{year}{1994}),
  \eprint{nucl-th/9406039}.

\bibitem[{\citenamefont{Wiringa et~al.}(1995)\citenamefont{Wiringa, Stoks, and
  Schiavilla}}]{Wiringa:1994wb}
\bibinfo{author}{\bibfnamefont{R.~B.} \bibnamefont{Wiringa}},
  \bibinfo{author}{\bibfnamefont{V.~G.~J.} \bibnamefont{Stoks}},
  \bibnamefont{and}
  \bibinfo{author}{\bibfnamefont{R.}~\bibnamefont{Schiavilla}},
  \bibinfo{journal}{Phys. Rev.} \textbf{\bibinfo{volume}{C51}},
  \bibinfo{pages}{38} (\bibinfo{year}{1995}), \eprint{nucl-th/9408016}.

\bibitem[{\citenamefont{Machleidt}(2001)}]{Machleidt:2000ge}
\bibinfo{author}{\bibfnamefont{R.}~\bibnamefont{Machleidt}},
  \bibinfo{journal}{Phys. Rev.} \textbf{\bibinfo{volume}{C63}},
  \bibinfo{pages}{024001} (\bibinfo{year}{2001}).

\bibitem[{\citenamefont{Gross and Stadler}(2008)}]{Gross:2008ps}
\bibinfo{author}{\bibfnamefont{F.}~\bibnamefont{Gross}} \bibnamefont{and}
  \bibinfo{author}{\bibfnamefont{A.}~\bibnamefont{Stadler}},
  \bibinfo{journal}{Phys.Rev.} \textbf{\bibinfo{volume}{C78}},
  \bibinfo{pages}{014005} (\bibinfo{year}{2008}), \eprint{0802.1552}.

\bibitem[{\citenamefont{Rentmeester et~al.}(1999)\citenamefont{Rentmeester,
  Timmermans, Friar, and de~Swart}}]{Rentmeester:1999vw}
\bibinfo{author}{\bibfnamefont{M.~C.~M.} \bibnamefont{Rentmeester}},
  \bibinfo{author}{\bibfnamefont{R.~G.~E.} \bibnamefont{Timmermans}},
  \bibinfo{author}{\bibfnamefont{J.~L.} \bibnamefont{Friar}}, \bibnamefont{and}
  \bibinfo{author}{\bibfnamefont{J.~J.} \bibnamefont{de~Swart}},
  \bibinfo{journal}{Phys. Rev. Lett.} \textbf{\bibinfo{volume}{82}},
  \bibinfo{pages}{4992} (\bibinfo{year}{1999}), \eprint{nucl-th/9901054}.


\bibitem[{\citenamefont{Rentmeester et~al.}(2003)\citenamefont{Rentmeester,
  Timmermans, and de~Swart}}]{Rentmeester:2003mf}
\bibinfo{author}{\bibfnamefont{M.}~\bibnamefont{Rentmeester}},
  \bibinfo{author}{\bibfnamefont{R.}~\bibnamefont{Timmermans}},
  \bibnamefont{and} \bibinfo{author}{\bibfnamefont{J.~J.}
  \bibnamefont{de~Swart}}, \bibinfo{journal}{Phys.Rev.}
  \textbf{\bibinfo{volume}{C67}}, \bibinfo{pages}{044001}
  (\bibinfo{year}{2003}), \eprint{nucl-th/0302080}.


\bibitem[{\citenamefont{Perez et~al.}(2013{\natexlab{a}})\citenamefont{Perez,
  Amaro, and Arriola}}]{Perez:2013mwa}
\bibinfo{author}{\bibfnamefont{R.~N.} \bibnamefont{Perez}},
  \bibinfo{author}{\bibfnamefont{J.}~\bibnamefont{Amaro}}, \bibnamefont{and}
  \bibinfo{author}{\bibfnamefont{E.~Ruiz} \bibnamefont{Arriola}},
  \bibinfo{journal}{Phys.Rev.} \textbf{\bibinfo{volume}{C88}},
  \bibinfo{pages}{024002} (\bibinfo{year}{2013}{\natexlab{a}}),
  \eprint{1304.0895}.

\bibitem[{\citenamefont{Perez et~al.}(2013{\natexlab{b}})\citenamefont{Perez,
  Amaro, and Arriola}}]{Perez:2013jpa}
\bibinfo{author}{\bibfnamefont{R.~N.} \bibnamefont{Perez}},
  \bibinfo{author}{\bibfnamefont{J.}~\bibnamefont{Amaro}}, \bibnamefont{and}
  \bibinfo{author}{\bibfnamefont{E.~R.} \bibnamefont{Arriola}}
  \bibinfo{journal}{Phys.Rev.} \textbf{\bibinfo{volume}{C88}},
  \bibinfo{pages}{064002} (\bibinfo{year}{2013},
 (\bibinfo{year}{2013}{\natexlab{b}}),\eprint{1310.2536}.

\bibitem[{\citenamefont{Aviles}(1972)}]{Aviles:1973ee}
\bibinfo{author}{\bibfnamefont{J.~B.} \bibnamefont{Aviles}},
  \bibinfo{journal}{Phys. Rev.} \textbf{\bibinfo{volume}{C6}},
  \bibinfo{pages}{1467} (\bibinfo{year}{1972}).

\bibitem[{\citenamefont{Entem et~al.}(2008)\citenamefont{Entem, Ruiz~Arriola,
  Pavon~Valderrama, and Machleidt}}]{Entem:2007jg}
\bibinfo{author}{\bibfnamefont{D.}~\bibnamefont{Entem}},
  \bibinfo{author}{\bibfnamefont{E.}~\bibnamefont{Ruiz~Arriola}},
  \bibinfo{author}{\bibfnamefont{M.}~\bibnamefont{Pavon~Valderrama}},
  \bibnamefont{and}
  \bibinfo{author}{\bibfnamefont{R.}~\bibnamefont{Machleidt}},
  \bibinfo{journal}{Phys.Rev.} \textbf{\bibinfo{volume}{C77}},
  \bibinfo{pages}{044006} (\bibinfo{year}{2008}), \eprint{0709.2770}.

\bibitem[{\citenamefont{Navarro~Perez
  et~al.}(2012{\natexlab{a}})\citenamefont{Navarro~Perez, Amaro, and
  Ruiz~Arriola}}]{NavarroPerez:2011fm}
\bibinfo{author}{\bibfnamefont{R.}~\bibnamefont{Navarro~Perez}},
  \bibinfo{author}{\bibfnamefont{J.}~\bibnamefont{Amaro}}, \bibnamefont{and}
  \bibinfo{author}{\bibfnamefont{E.}~\bibnamefont{Ruiz~Arriola}},
  \bibinfo{journal}{Prog.Part.Nucl.Phys.} \textbf{\bibinfo{volume}{67}},
  \bibinfo{pages}{359} (\bibinfo{year}{2012}{\natexlab{a}}),
  \eprint{1111.4328}.




\bibitem[{\citenamefont{Navarro~Perez
  et~al.}(2012{\natexlab{b}})\citenamefont{Navarro~Perez, Amaro, and
  Ruiz~Arriola}}]{NavarroPerez:2012qf}
\bibinfo{author}{\bibfnamefont{R.}~\bibnamefont{Navarro~Perez}},
  \bibinfo{author}{\bibfnamefont{J.}~\bibnamefont{Amaro}}, \bibnamefont{and}
  \bibinfo{author}{\bibfnamefont{E.}~\bibnamefont{Ruiz~Arriola}}
  \bibinfo{journal}{Phys.Lett.} \textbf{\bibinfo{volume}{B724}},
  \bibinfo{pages}{138} (\bibinfo{year}{2013}{\natexlab{a}}), \eprint{1202.2689}.

\bibitem[{\citenamefont{Navarro~Perez et~al.}(2013)\citenamefont{Navarro~Perez,
  Amaro, and Ruiz~Arriola}}]{Perez:2013za}
\bibinfo{author}{\bibfnamefont{R.}~\bibnamefont{Navarro~Perez}},
  \bibinfo{author}{\bibfnamefont{J.}~\bibnamefont{Amaro}}, \bibnamefont{and}
  \bibinfo{author}{\bibfnamefont{E.}~\bibnamefont{Ruiz~Arriola}},
  \bibinfo{journal}{PoS} \textbf{\bibinfo{volume}{CD12}}, \bibinfo{pages}{104}
  (\bibinfo{year}{2013}), \eprint{1301.6949}.

\bibitem{Perez:2013cza} 
  R.~N.~Perez, J.~E.~Amaro and E.~R.~Arriola,
  arXiv:1310.8167 [nucl-th].


\bibitem{Stoks:1992ja} 
  V.~G.~J.~Stoks, R.~Timmermans and J.~J.~de Swart,
  Phys.\ Rev.\ C {\bf 47}, 512 (1993)
  [nucl-th/9211007].


\bibitem{deSwart:1997ep} 
  J.~J.~de Swart, M.~C.~M.~Rentmeester and R.~G.~E.~Timmermans,
  PiN Newslett.\  {\bf 13}, 96 (1997)
  [nucl-th/9802084].



\bibitem[{\citenamefont{Calle~Cordon et~al.}(2012)\citenamefont{Calle~Cordon,
  Pavon~Valderrama, and Ruiz~Arriola}}]{CalleCordon:2010sq}
\bibinfo{author}{\bibfnamefont{A.}~\bibnamefont{Calle~Cordon}},
  \bibinfo{author}{\bibfnamefont{M.}~\bibnamefont{Pavon~Valderrama}},
  \bibnamefont{and}
  \bibinfo{author}{\bibfnamefont{E.}~\bibnamefont{Ruiz~Arriola}},
  \bibinfo{journal}{Phys.Rev.} \textbf{\bibinfo{volume}{C85}},
  \bibinfo{pages}{024002} (\bibinfo{year}{2012}), \eprint{1010.1728}.

\bibitem[{\citenamefont{Miller et~al.}(2006)\citenamefont{Miller, Opper, and
  Stephenson}}]{Miller:2006tv}
\bibinfo{author}{\bibfnamefont{G.~A.} \bibnamefont{Miller}},
  \bibinfo{author}{\bibfnamefont{A.~K.} \bibnamefont{Opper}}, \bibnamefont{and}
  \bibinfo{author}{\bibfnamefont{E.~J.} \bibnamefont{Stephenson}},
  \bibinfo{journal}{Ann.Rev.Nucl.Part.Sci.} \textbf{\bibinfo{volume}{56}},
  \bibinfo{pages}{253} (\bibinfo{year}{2006}), \eprint{nucl-ex/0602021}.

\bibitem[{\citenamefont{van Kolck et~al.}(1998)\citenamefont{van Kolck,
  Rentmeester, Friar, Goldman, and de~Swart}}]{vanKolck:1997fu}
\bibinfo{author}{\bibfnamefont{U.}~\bibnamefont{van Kolck}},
  \bibinfo{author}{\bibfnamefont{M.}~\bibnamefont{Rentmeester}},
  \bibinfo{author}{\bibfnamefont{J.~L.} \bibnamefont{Friar}},
  \bibinfo{author}{\bibfnamefont{J.~T.} \bibnamefont{Goldman}},
  \bibnamefont{and} \bibinfo{author}{\bibfnamefont{J.}~\bibnamefont{de~Swart}},
  \bibinfo{journal}{Phys.Rev.Lett.} \textbf{\bibinfo{volume}{80}},
  \bibinfo{pages}{4386} (\bibinfo{year}{1998}), \eprint{nucl-th/9710067}.

\bibitem[{\citenamefont{Alarcon et~al.}(2013)\citenamefont{Alarcon,
  Martin~Camalich, and Oller}}]{Alarcon:2012kn}
\bibinfo{author}{\bibfnamefont{J.}~\bibnamefont{Alarcon}},
  \bibinfo{author}{\bibfnamefont{J.}~\bibnamefont{Martin~Camalich}},
  \bibnamefont{and} \bibinfo{author}{\bibfnamefont{J.}~\bibnamefont{Oller}},
  \bibinfo{journal}{Annals Phys.} \textbf{\bibinfo{volume}{336}},
  \bibinfo{pages}{413} (\bibinfo{year}{2013}), \eprint{1210.4450}.


\bibitem[{\citenamefont{Ekstr{\"o}m et~al.}(2013)\citenamefont{Ekstr{\"o}m,
  Baardsen, Forssén, Hagen, Hjorth-Jensen et~al.}}]{Ekstrom:2013kea}
\bibinfo{author}{\bibfnamefont{A.}~\bibnamefont{Ekstr{\"o}m}},
  \bibinfo{author}{\bibfnamefont{G.}~\bibnamefont{Baardsen}},
  \bibinfo{author}{\bibfnamefont{C.}~\bibnamefont{Forssén}},
  \bibinfo{author}{\bibfnamefont{G.}~\bibnamefont{Hagen}},
  \bibinfo{author}{\bibfnamefont{M.}~\bibnamefont{Hjorth-Jensen}},
  \bibnamefont{et~al.}, \bibinfo{journal}{Phys.Rev.Lett.}
  \textbf{\bibinfo{volume}{110}}, \bibinfo{pages}{192502}
  (\bibinfo{year}{2013}), \eprint{1303.4674}.

\bibitem{Amaro:2013zka}
  J.~E.~Amaro, R.~N.~Perez and E.~R.~Arriola,
  arXiv:1310.7456 [nucl-th].


\bibitem[{\citenamefont{Pavon~Valderrama and
  Ruiz~Arriola}(2006)}]{PavonValderrama:2005wv}
\bibinfo{author}{\bibfnamefont{M.}~\bibnamefont{Pavon~Valderrama}}
  \bibnamefont{and}
  \bibinfo{author}{\bibfnamefont{E.}~\bibnamefont{Ruiz~Arriola}},
  \bibinfo{journal}{Phys.Rev.} \textbf{\bibinfo{volume}{C74}},
  \bibinfo{pages}{054001} (\bibinfo{year}{2006}), \eprint{nucl-th/0506047}.


\bibitem[{\citenamefont{Entem and Machleidt}(2003)}]{Entem:2003ft}
\bibinfo{author}{\bibfnamefont{D.}~\bibnamefont{Entem}} \bibnamefont{and}
  \bibinfo{author}{\bibfnamefont{R.}~\bibnamefont{Machleidt}},
  \bibinfo{journal}{Phys.Rev.} \textbf{\bibinfo{volume}{C68}},
  \bibinfo{pages}{041001} (\bibinfo{year}{2003}).



\bibitem[{\citenamefont{Buettiker and Meissner}(2000)}]{Buettiker:1999ap}
\bibinfo{author}{\bibfnamefont{P.}~\bibnamefont{Buettiker}} \bibnamefont{and}
  \bibinfo{author}{\bibfnamefont{U.-G.} \bibnamefont{Meissner}},
  \bibinfo{journal}{Nucl.Phys.} \textbf{\bibinfo{volume}{A668}},
  \bibinfo{pages}{97} (\bibinfo{year}{2000}).


\bibitem[{\citenamefont{Navarro~Perez
  et~al.}(2012{\natexlab{c}})\citenamefont{Navarro~Perez, Amaro, and
  Ruiz~Arriola}}]{NavarroPerez:2012vr}
\bibinfo{author}{\bibfnamefont{R.}~\bibnamefont{Navarro~Perez}},
  \bibinfo{author}{\bibfnamefont{J.}~\bibnamefont{Amaro}}, \bibnamefont{and}
  \bibinfo{author}{\bibfnamefont{E.}~\bibnamefont{Ruiz~Arriola}}
  (\bibinfo{year}{2012}{\natexlab{c}}), \eprint{1202.6624}.

\bibitem[{\citenamefont{Gilman and Gross}(2002)}]{Gilman:2001yh}
\bibinfo{author}{\bibfnamefont{R.~A.} \bibnamefont{Gilman}} \bibnamefont{and}
  \bibinfo{author}{\bibfnamefont{F.}~\bibnamefont{Gross}},
  \bibinfo{journal}{J.Phys.} \textbf{\bibinfo{volume}{G28}},
  \bibinfo{pages}{R37} (\bibinfo{year}{2002}), \eprint{nucl-th/0111015}.


\bibitem[{\citenamefont{Leun and Alderliesten}(1982)}]{VanDerLeun1982261}
\bibinfo{author}{\bibfnamefont{C.~V.~D.} \bibnamefont{Leun}} \bibnamefont{and}
  \bibinfo{author}{\bibfnamefont{C.}~\bibnamefont{Alderliesten}},
  \bibinfo{journal}{Nucl. Phys.} \textbf{\bibinfo{volume}{A380}},
  \bibinfo{pages}{261 } (\bibinfo{year}{1982}).

\bibitem[{\citenamefont{Borbély et~al.}(1985)\citenamefont{Borbély,
  Grüebler, König, Schmelzbach, and Mukhamedzhanov}}]{Borbély198517}
\bibinfo{author}{\bibfnamefont{I.}~\bibnamefont{Borbély}},
  \bibinfo{author}{\bibfnamefont{W.}~\bibnamefont{Grüebler}},
  \bibinfo{author}{\bibfnamefont{V.}~\bibnamefont{König}},
  \bibinfo{author}{\bibfnamefont{P.~A.} \bibnamefont{Schmelzbach}},
  \bibnamefont{and} \bibinfo{author}{\bibfnamefont{A.~M.}
  \bibnamefont{Mukhamedzhanov}}, \bibinfo{journal}{Phys. Lett.}
  \textbf{\bibinfo{volume}{160B}}, \bibinfo{pages}{17 } (\bibinfo{year}{1985}).

\bibitem[{\citenamefont{Rodning and Knutson}(1990)}]{Rodning:1990zz}
\bibinfo{author}{\bibfnamefont{N.~L.} \bibnamefont{Rodning}} \bibnamefont{and}
  \bibinfo{author}{\bibfnamefont{L.~D.} \bibnamefont{Knutson}},
  \bibinfo{journal}{Phys. Rev.} \textbf{\bibinfo{volume}{C41}},
  \bibinfo{pages}{898} (\bibinfo{year}{1990}).

\bibitem[{\citenamefont{Klarsfeld et~al.}(1986)\citenamefont{Klarsfeld,
  Martorell, Oteo, Nishimura, and Sprung}}]{Klarsfeld1986373}
\bibinfo{author}{\bibfnamefont{S.}~\bibnamefont{Klarsfeld}},
  \bibinfo{author}{\bibfnamefont{J.}~\bibnamefont{Martorell}},
  \bibinfo{author}{\bibfnamefont{J.~A.} \bibnamefont{Oteo}},
  \bibinfo{author}{\bibfnamefont{M.}~\bibnamefont{Nishimura}},
  \bibnamefont{and} \bibinfo{author}{\bibfnamefont{D.~W.~L.}
  \bibnamefont{Sprung}}, \bibinfo{journal}{Nucl. Phys.}
  \textbf{\bibinfo{volume}{A456}}, \bibinfo{pages}{373 }
  (\bibinfo{year}{1986}).

\bibitem[{\citenamefont{Bishop and Cheung}(1979)}]{Bishop:1979zz}
\bibinfo{author}{\bibfnamefont{D.~M.} \bibnamefont{Bishop}} \bibnamefont{and}
  \bibinfo{author}{\bibfnamefont{L.~M.} \bibnamefont{Cheung}},
  \bibinfo{journal}{Phys. Rev.} \textbf{\bibinfo{volume}{A20}},
  \bibinfo{pages}{381} (\bibinfo{year}{1979}).

\bibitem[{\citenamefont{de~Swart et~al.}(1995)\citenamefont{de~Swart,
  Terheggen, and Stoks}}]{deSwart:1995ui}
\bibinfo{author}{\bibfnamefont{J.~J.} \bibnamefont{de~Swart}},
  \bibinfo{author}{\bibfnamefont{C.~P.~F.} \bibnamefont{Terheggen}},
  \bibnamefont{and} \bibinfo{author}{\bibfnamefont{V.~G.~J.}
  \bibnamefont{Stoks}} (\bibinfo{year}{1995}), \eprint{nucl-th/9509032}.


\end{thebibliography}

\end{document}